\documentclass{article}

\usepackage{PRIMEarxiv}

\usepackage{amsmath}
\usepackage{lineno,hyperref}
\usepackage{amssymb}
\usepackage{amsthm}
\usepackage{epsfig}
\usepackage{graphicx}
\usepackage{graphics}
\usepackage{float}
\usepackage{subcaption}
\usepackage{multirow}
\usepackage{xcolor}
\usepackage{comment}
\usepackage{euscript}
\usepackage[normalem]{ulem} 
\usepackage{makeidx}
\usepackage{multirow}
\usepackage{xspace}
\usepackage{wrapfig}
\usepackage{bm}
\usepackage{mathrsfs}
\usepackage{contour}
\usepackage{breqn}
\usepackage{algorithm}
\usepackage{footnote}

\usepackage[utf8]{inputenc}
\usepackage{amssymb,amsthm,amsmath,xcolor,mathrsfs}
\usepackage{lineno}
\usepackage{graphicx}
\usepackage{graphics}
\usepackage{float}
\usepackage{caption}
\usepackage{subcaption}
\usepackage{multirow}
\usepackage{color}
\usepackage{comment}
\usepackage{euscript}
\usepackage{hhline}
\usepackage{makeidx}
\usepackage{multirow}
\usepackage{xspace}
\usepackage{wrapfig}
\usepackage{bm}
\usepackage{soul}
\usepackage{contour}
\usepackage{breqn}
\usepackage{algorithm}
\usepackage{footnote}
\usepackage{tikz-cd}
\usepackage{tikz}
\usepackage{indentfirst,csquotes}
\usepackage[normalem]{ulem} 
\usepackage{breqn}
\usepackage{algorithm}
\usepackage{footnote}
\makesavenoteenv{tabular}
\makesavenoteenv{table}
\usepackage{algpseudocode}

\usepackage{xcolor,paralist,hyperref,titlesec,fancyhdr,etoolbox}
\usepackage{blindtext}

\makeatletter
\def\BState{\State\hskip-\ALG@thistlm}
\makeatother
\algnewcommand\algorithmicinput{\textbf{Input:}}
\algnewcommand\INPUT{\item[\algorithmicinput]}
\makeatother
\algnewcommand\algorithmicoutput{\textbf{Output:}}
\algnewcommand\OUTPUT{\item[\algorithmicoutput]}

\modulolinenumbers[1]

\usetikzlibrary{shapes,arrows}

\newcommand{\lbl}[1]{\label{#1}}

\newcommand{\nc}{N_{\text{c}}}
\newcommand{\eps}{\epsilon}
\newcommand{\vc}{\mathbf}

\definecolor{darkred}{rgb}{1, 0.1, 0.3}
\definecolor{darkblue}{rgb}{0.1, 0.1, 1}
\definecolor{darkgreen}{rgb}{0,0.6,0.5}

\makeatletter
\newenvironment{breakablealgorithm}
  {
   \begin{center}
     \refstepcounter{algorithm}
     \hrule height.8pt depth0pt \kern2pt
     \renewcommand{\caption}[2][\relax]{
       {\raggedright\textbf{\fname@algorithm~\thealgorithm} ##2\par}%
       \ifx\relax##1\relax 
         \addcontentsline{loa}{algorithm}{\protect\numberline{\thealgorithm}##2}%
       \else 
         \addcontentsline{loa}{algorithm}{\protect\numberline{\thealgorithm}##1}%
       \fi
       \kern2pt\hrule\kern2pt
     }
  }{
     \kern2pt\hrule\relax
   \end{center}
  }
\makeatother

\title{A parallel-in-time method based on the Parareal algorithm and High-Order Dynamic Mode Decomposition with applications to fluid simulations
}

\author{
 Weifan Liu \\
 College of Science\\
  Beijing Forestry University \\
  Beijing, China, 100083\\
  \texttt{weifanliu@bjfu.edu.cn} \\
}

\begin{document}
\maketitle

\begin{abstract}
The high cost of sequential time integration is one major constraint that limits the speedup of a time-parallel algorithm like the Parareal algorithm due to the difficulty of coarsening time steps in a stiff numerical problem. To address this challenge, we develop a parallel-in-time approach based on the Parareal algorithm, in which we construct a novel coarse solver using a data-driven method based on Dynamic Mode Decomposition in place of a classic time marching scheme. The proposed solver computes an approximation of the solution using two numerical schemes of different accuracies in parallel, and apply High-Order Dynamic Mode Decomposition (HODMD) to reduce the cost of sequential computations. Compared to the original Parareal algorithm, the proposed approach allows for the construction of low-cost coarse solvers for many complicated stiff problems. We demonstrate through several numerical examples in fluid dynamics that the proposed method can effectively reduce the serial computation cost and improve the parallel speedup of long-time simulations which are hard to accelerate using the original Parareal algorithm.
\end{abstract}


\section{Introduction}
Traditional time-stepping methods for solving time-dependent problems are inherently sequential, which can be a bottleneck for some large-scale simulations. Parallel-in-time methods are a class of algorithms designed to accelerate the numerical solution of time-dependent problems by exploiting parallelism in the time dimension. These methods aim to overcome this limitation by distributing the computation across multiple processors in the time dimension. The methods have gained increasing attention in recent years due to their potential in reducing computation time for large-scale simulations, especially when the parallel speedup of spatial parallelization saturates. Gander classified the parallel-in-time methods into four main classes, including multiple shooting, domain decomposition and waveform relaxation-based methods, space-time multigrid methods and direct time parallel methods \cite{gander201550}. The Parareal algorithm is one commonly used parallel-in-time method, developed by Lions et al. \cite{lionsparareal}. The algorithm employs a fine and a coarse solver, and iteratively corrects the solution by alternating between the fine and the coarse solver. Since its introduction, the algorithm has been widely used in accelerating fluid problems \cite{eghbal2017acceleration, liu2022parallel, steinstraesser2024parallel}, neural network-based partial differential equations (PDEs) solvers \cite{meng2020ppinn}, linear switched systems \cite{li2022study}, Hamiltonian systems \cite{gander2014analysis}, etc. Multigrid-in-Time (MGRIT), an extension of the multigrid method to the time dimension, is one other non-intrusive parallel-in-time approach that interprets the parareal time integration as a two-level reduction method \cite{friedhoff2012multigrid}. MGRIT has been used to solve non-linear hyberbolic equations \cite{Danieli2021MultigridRI}, space-fractional diffusion equation \cite{yue2019parallel}, and Stokes and Oseen problems \cite{yoda2024coarse}, etc. Other parallel-in-time approaches include wavefront relaxation \cite{gear1991waveform}, Parallel Full Approximation Scheme in Space and Time (PFASST) \cite{emmett2012toward}, and direct parallelization of existing time-stepping methods \cite{miranker1967parallel}.

The high cost of sequential computation of the Parareal algorithm can significantly hinder the parallel speedup and efficiency of the algorithm. To address this issue, many variants of the Parareal algorithm have been developed. Legoll et al. proposed a Micro-Macro Parareal algorithm, which assumes the variables can be separated into slow and fast components and uses a coarse propagator to approximate the macroscopic model of the slow components \cite{MicroMacroParareal}. The algorithm possesses a low-cost coarse solver due to the elimination of the fast variables and the possibility of using a much larger time step. GParareal is another Parareal-based algorithm which replaces the correction term in the original Parareal algorithm using a Gaussian process emulator \cite{2023GParareal}. The algorithm uses data from prior iterations to train the emulator, and has been shown to achieve improved convergence, runtime and accuracy on low-dimensional ODEs. The accuracy of the emulator, however, can be influenced by the data points of the generated solution. With the development of deep learning methods in numerical PDEs, Parareal with a Physics-Informed Neural Network (PINN)-based coarse propagator is developed \cite{PararealPINN}, where a PINN-based trained neural network model is used as the coarse solver to reduce the sequential cost of the Parareal algorithm and maximize the utilization of heterogeneous architectures by combining the use of CPUs and GPUs. 

In the abovementioned methods like the regular Parareal, MGRIT or Micro-Macro Parareal, the use of the coarse grid in the time dimension is essential to achieve high speedup and parallel efficiency of parallel computations. However, in the case of long-time simulations of stiff problems, one of the main challenges and bottlenecks is that the stiffness of the problem does not allow for a drastic relaxation on the time step in a classic time-marching numerical scheme. The upper bound on the speedup of Parareal depends largely on the computational cost of the coarse solver, which is sequential in nature. As stiff problems have demanding restrictions on the time step, the coarse solver based on a natural coarsening of time discretization is thus often prohibitive, which significantly limits the speedup of a time parallel algorithm like the Parareal algorithm. To resolve the difficulty of using larger time steps and to improve the parallel speedup of the algorithm in such scenarios, we propose a novel parallel-in-time approach based on the Parareal algorithm and High-Order Dynamic Mode Decomposition (HODMD). Specifically, we propose to construct a coarse solver by applying two numerical schemes of high and low accuracy respectively in parallel. In the more accurate and more expensive numerical scheme, we compute the solution only on a subset of each partitioned time interval, on which we also compute the difference between the solutions given by the two schemes. HODMD \cite{le2017higher} is then applied to the difference to predict the solution at the endpoint of each interval. We combine this prediction with the solution given by the less accurate numerical scheme to obtain an approximation of the coarse propagator of the Parareal algorithm. The proposed coarse solver provides a new approach for constructing coarse approximations to stiff differential equations with a reduced computation cost compared to the classic coarse solver. The proposed solver, when embedded inside of the Parareal iteration, reduces the sequential cost of the Parareal algorithm and is capable of achieving higher speedup in some stiff problems which are difficult to accelerate using the traditional Parareal algorithm with classic solvers. 

In this work, we focus on the numerical simulations of some complicated fluid dynamics problems with high time step restrictions, including the motion of dynamic structures immersed in incompressible viscous fluid and the dewetting of thin liquid film droplets on patterned substrates. Parallel-in-time methods have been previously investigated for the temporal parallelization of fluid simulations \cite{liu2022parallel, 2014ParallelRandles, 2021MultiscaleZebra, Araz2017Acceleration,2010Modified}. However, in some fluid simulations like \cite{liu2022parallel}, the stability requirement of the problem due to the numerical regularization parameter and the physical parameters involved pose challenges in constructing cheap coarse propagators. In this work, we apply the proposed novel coarse solver to such biofluid problems where the time step is restrictive, and show through several numerical examples that the proposed approach effectively reduces the total computation time and improves the parallel efficiency of some long-time numerical simulations compared to the original Parareal method.

The rest of the paper is organized as follows. In Section \ref{sec:parareal}, we review the Parareal algorithm. In Section \ref{sec:dmdintro}, we review the concepts and variants of DMD including HODMD. In Section \ref{sec:parareal-hodmd}, we describe the construction of the novel coarse solver and how we incorporate HODMD into the coarse solver in the Parareal iterations. In Section \ref{sec:examples}, we apply the proposed algorithm to various numerical experiments, including the simulation of rod-like microswimmer, elastic sphere, and dewetting thin liquid films. We demonstrate the efficiency of the proposed method by presenting and analyzing the accuracy, speedup, scaling of the proposed method applied to the above numerical examples. We conclude the article in Section \ref{sec:conclude}.

\section{Parareal Algorithm}\label{sec:parareal}
The Parareal algorithm is a parallel-in-time algorithm introduced in \cite{lionsparareal}. The algorithm is designed to solve time-dependent problems by parallelizing the time domain of the solution. Consider an Initial Value Problem (IVP) of the following form:
\begin{linenomath*}
\begin{equation}\label{eq:ode}
\frac{dx}{dt}=u(t,x)~~\text{for}~~t\in(0,T]~~\text{and}~~{x}(0)={x}_0.
\end{equation}
\end{linenomath*}
Assume that there are $\nc$ cores. The Parareal algorithm divides the time domain $[0,T]$ uniformly into $\nc$ subintervals of length $\Delta T=T/\nc$ and let $T_n=n\Delta T$ for $n=0,~1,~2,~\cdots,~\nc$. The algorithm parallelizes the computation by iteratively alternating between serial calculations based on a coarse solver and parallel calculations based on a fine solver. Compared to the coarse solver, the fine solver is a more accurate solver, which is typically a high-order numerical scheme with finer grid. Let $\mathcal{G}(T_n,T_{n-1}, x_{n-1}^k)$ and $\mathcal{F}(T_n,T_{n-1}, X_{n-1}^k)$ denote the coarse solver and the fine solver used to solve the IVP:
\begin{linenomath*}
\begin{equation}\label{eq:oden}
\frac{dx}{dt} = u(t, x), \quad x(T_{n-1}) = x_{n-1}^k,
\end{equation}
\end{linenomath*}
for $t\in(T_{n-1}, T_n]$, $n=1,2,\cdots,\nc$. The main ingredients of a regular Parareal algorithm for solving Equation \eqref{eq:ode} are given below:
\begin{enumerate}
\item For $n=k,~k+1,~k+2,~\cdots,~\nc$, compute in serial 
\begin{linenomath*}
\begin{equation}\label{eq:initial}
X^0_{n}=\mathcal{G}(T_{n},T_{n-1},X^0_{n-1}), \quad X_0^0=x_0.
\end{equation}
\end{linenomath*}
\item For $n=k,~k+1,~k+2,~\cdots,~\nc$, compute in parallel
\begin{linenomath*}
\begin{equation}\label{eq:intermediate}
{X_n^k}^{\prime}=\mathcal{F}\left(T_{n},T_{n-1},X^{k-1}_{n-1}\right) .
\end{equation}
\end{linenomath*}
\item Let $X_n^k=X_n^{k-1}$ for $n=1,~2,~\cdots,~k-1$ and $X_k^k = {X_k^k}^{\prime}$. For $n=k+1,~k+2,~\cdots,~\nc$, correct ${X_n^k}^{\prime}$ by applying the coarse solver sequentially as follows:
\begin{linenomath*}
\begin{equation}\label{eq:serial_correct}
X^{k}_{n}={X_n^k}^{\prime}+\underbrace{\mathcal{G}\left(T_{n},T_{n-1},X^{k}_{n-1}\right)-\mathcal{G}\left(T_{n},T_{n-1},X^{k-1}_{n-1}\right)}_{\text{\normalsize correction term}},
\end{equation}
\end{linenomath*}
\end{enumerate}
where $\mathcal{G}\left(T_{n},T_{n-1},X^{k-1}_{n-1}\right)$ in Eq. (\ref{eq:serial_correct}) has been computed in the previous iteration.

\section{Dynamic Mode Decomposition}\label{sec:dmdintro}
Dynamic Mode Decomposition (DMD) is a dimensionality reduction technique for sequential data, and has been used for analyzing dynamical systems, particularly those that exhibit complex behavior. Based on the theory of Koopman analysis, DMD seeks to decompose the dynamics of a system into a set of modes. With its roots in the field of fluid dynamics, the method has since been applied to a wide range of complicated problems in power systems \cite{Barocio2015ADM}, finance\cite{Kuttichira2017StockPP}, medical image analysis\cite{mriDMD}, etc.

DMD operates on the principle of decomposing a linear approximation of a nonlinear system's dynamics. Given a sequence of snapshots from a dynamical system, the standard DMD seeks to find a matrix that best represents the evolution of the system from one snapshot to the next. Assume ${X}=[\vc{x}_0, \vc{x}_1,\cdots, \vc{x}_{m-1}]$ and ${X}'=[\vc{x}_1, \vc{x}_2,\cdots, \vc{x}_{m}]$ are two matrices, each consisting of $m$ snapshots from the sequential data $\{\vc{x}_i\}_{i=0}^{m}$. The idea of DMD is to approximate the time evolution of the sequential data by seeking a linear system such that ${X}'=A{X}$, i.e. $\vc{x}_{k+1}=A\vc{x}_k$. The best fit $A$ is given by $A={X}'{X}^{\dagger}$, where $X^{\dagger}$ is the Moore–Penrose pseudoinverse of ${X}$.

To find ${X}'{X}^{\dagger}$, the algorithm computes a Singular Value Decomposition (SVD) of ${X}$, and constructs $\tilde{A}_r$, an approximation of $A$ by using the SVD decomposition that preserves only the $r$ largest single values for some $r\ll m$, i.e.
\begin{subequations}
\begin{align}
X & \approx  U_r\Sigma_rV_r^{*}, \\    
\tilde{A}_r &= U_r^{*}AU_r,
\end{align}
\end{subequations}
where $^*$ denotes the conjugate transpose of the matrix. The eigen-decomposition of $\tilde A=W\Lambda W^{-1}$ is then applied to obtain the first $r$ eigenvalues and their corresponding eigenvectors of $\tilde A$, where $\Lambda$ is a diagonal matrix with eigenvalues of $\tilde A$ and $W$ is a matrix consisting of the corresponding eigenvectors. The first $r$ eigenvalues of $A$ are the same as that of $\tilde A$. The corresponding eigenvectors of $A$ can be obtained via the reconstruction $\Phi=X'V\Sigma^{-1}W$. The prediction of the time evolution of $x(t)$ is then given by
\begin{equation}
\vc{x}(t)\approx \Phi e^{\Lambda t}\vc{b},
\end{equation}
where $\vc b$ denotes the coefficients of the initial condition projected onto the eigenmodes \cite{2015OnTu}.

Since its introduction, many variants of the standard DMD algorithm has been developed. These variants aim to improve the accuracy, robustness and efficiency of the DMD algorithm. Examples of such improvements include exact DMD \cite{tu2013dynamic}, multiresolution DMD \cite{kutz2016multiresolution}, DMD with control \cite{brunton2016koopman,proctor2016dynamic}, high-order DMD (HODMD) \cite{le2017higher}, etc. In this work, we focus on HODMD and incorporate it in the initial serial sweep and serial correction of the Parareal algorthm to estimate the serial solution. The HODMD algorithm is a generalization of the standard DMD algorithm with time-lagged snapshots. The algorithm has been shown to calculate the decomposition of modes in a more robust and accurate way and has been applied to a variety of fluid problems \cite{2022Dynamic,2020Dynamic,2018Dynamic,2010Dynamic}. We briefly review the HODMD algorithm. Let $x_k$ denote the data at the $k$th snapshot, for $k=1,2,\cdots K.$ The standard DMD assumes $x_{k+1}=Ax_k, ~ k=1,2,\cdots, K-1$. For a higher-order estimate, the high-order DMD proposes to use 
\begin{equation}
x_{k+d} = A_1x_k +A_2x_{k+1}+\cdots+A_dx_{k+d-1}, ~ k = 1, \cdots, K-d,
\end{equation}
where $d$ is a tunable parameter that can be chosen based on the desired order. For implementation, the modification can be achieved by replacing the original snapshot $x_k$ with $\tilde{x}_k$ with the modified Koopman matrix $\tilde A$, given by
\begin{equation}
\tilde{x}_k = \begin{bmatrix}
x_k\\
x_{k+1}\\
\vdots\\
x_{k+d-1},
\end{bmatrix}\quad 
\tilde{R}=\begin{bmatrix}
0, & I, & 0, & \cdots, & 0, & 0\\ 
0, & 0, & I, & \cdots, & 0, & 0\\
\cdots, & \cdots, & \cdots, & \cdots, & \cdots, & \cdots\\
0, & 0, & 0, & \cdots, & I, & 0\\
A_1, & A_2, & A_3, & \cdots, & A_{d-1}, & A_d
\end{bmatrix}.
\end{equation}

\section{Novel Parareal algorithm based on High-Order Dynamic Mode Decomposition}\label{sec:parareal-hodmd}
One major issue that limits the speedup of the Parareal algorithm is the high cost of the sequential calculation involved in the coarse solver. However, for complicated problems with high stability requirements, it is often prohibitive to build coarse solvers by directly coarsening the time mesh. To this end, we present an improved Parareal algorithm with a novel coarse solver, which produces coarse approximations of such problems by employing HODMD. We propose modifications to the two major serial calculations involved in the Parareal algorithm, i.e. the initial coarse serial sweep and the serial correction sweep during each Parareal iteration.
\subsection{Initial serial sweep}
To apply HODMD in the initial serial calculation, we first construct two coarse solvers $\mathcal{G}_1$ and $\mathcal{G}_2$, where $\mathcal{G}_2$ is a coarser solver than $\mathcal{G}_1$ in time and/or spatial dimension. For simplicity, we illustrate the idea of the algorithm using a one-dimensional problem. Suppose $X(s,t)$ is the solution to a time-dependent partial differential equation (PDE). Let $\tau_c^{(1)}$ and $\sigma_1^{(1)}$ denote the time step and spatial discretization space between each pair of consecutive grid points of $\mathcal{G}_1$ respectively, and $\tau_c^{(2)}$ and $\sigma_c^{(2)}$ the time step and discretization spacing of $\mathcal{G}_2$, where $\sigma_c^{(2)}\geq\sigma_c^{(1)}, ~\tau_c^{(2)} = r_q\tau_c^{(1)}$ for some integer $r_q\geq 1$. Let $K_t<\nc$ be an integer less than the total number of partitioned subintervals $\nc$. In the initial serial sweep, instead of sequentially computing $X_n^0=\mathcal{G}(T_n, T_{n-1}, X_{n-1}^0)$ for $n=1, 2, \cdots, \nc$ on a single core as in the classic Parareal algorithm, perform the following two computations on two cores in parallel:
\begin{align}
{\rm Core}~\#~0: ~ & {X}_{j}^0=\mathcal{G}_1(T_j, T_{j-1}, {X}_{j-1}^0), ~ j = 1,2,\cdots, K_t, \label{eq:core1}\\
{\rm Core}~\#~1: ~ & {X}_{j}^0=\mathcal{G}_2(T_j, T_{j-1}, {X}_{j-1}^0), ~ j = 1,2,\cdots,\nc. \label{eq:core2} 
\end{align}
That is, we calculate $\mathcal{G}_1(T_j, T_{j-1}, X_{j-1}^0)$, the more accurate solver of the two, on a subset of the time domain, and calculate $\mathcal{G}_2(T_j, T_{j-1}, X_{j-1}^0)$, the coarser and cheaper solver of the two on the entire time domain. Assume the parameters and the scheme of $\mathcal{G}_1$ and $\mathcal{G}_2$ are chosen so that the computation time of Equation \eqref{eq:core2} does not exceed that of Equation \eqref{eq:core1}, in which case, the time cost of the initial serial sweep is solely determined by the parameters of $\mathcal{G}_1$ and $K_t$, without considering the parallel overhead.

In the initial sweep, for $j = 1,2,\cdots, K_t$, we simply take $X_j^0=\mathcal{G}_1(T_j,T_{j-1},X_{j-1}^0)$, the more accurate of the two. For $K_t<j\leq\nc$, we calculate a ``correction'' to $\mathcal{G}_2(T_j, T_{j-1}, X_{j-1}^0)$ by using a data-driven prediction based on HODMD and the readily available solutions given by $\mathcal{G}_1$. To take advantage of the both solvers calculated in Equation \eqref{eq:core1} and \eqref{eq:core2}, we calculate the difference in the solutions given by $\mathcal{G}_1$ and $\mathcal{G}_2$ on the first $K_t$ sub-intervals, and construct a snapshot matrix using the evolution of the difference data. We apply HODMD to the difference snapshot matrix and use the HODMD prediction to estimate a ``correction'' to $\mathcal{G}_2(T_j, T_{j-1}, X_{j-1}^0)$ for $j=K_t+1,\cdots, \nc$. We formulate the process below. Let $X^{0}_{j,1}$ denote $\mathcal{G}_1(T_0+j\tau_c^{(1)}, T_0, X(s,0))$ and $X^{0}_{j,2}$ denote $\mathcal{G}_2(T_0+j\tau_c^{(1)}, T_0, X(s,0))$ respectively, i.e. the solutions computed by two numerical schemes of different costs and accuracies in parallel in the initial serial sweep. Once they are computed, construct two matrices $\Phi_1$ and $\Phi_2$ as follows:
\begin{align}
\label{eq:Phi1_defn}\Phi_1 & = \begin{bmatrix} X^0_{0,1}, X^0_{1,1}, \cdots, X^0_{p, 1}\end{bmatrix},\\
\label{eq:Phi2_defn}\Phi_2 & = \begin{bmatrix} \mathcal{I}_{\sigma^{(2)}_c}^{\sigma^{(1)}_c}X^0_{0,2}, \mathcal{I}_{\sigma^{(2)}_c}^{\sigma^{(1)}_c}X^0_{1,2}, \cdots, \mathcal{I}_{\sigma^{(2)}_c}^{\sigma^{(1)}_c}X^0_{p,2}\end{bmatrix},
\end{align}
where $p=N_{q1}\cdot K_t$, $N_{q1}$ is the number of coarse time steps computed by $\mathcal{G}_1$ on each subinterval of length $\Delta T=T/\nc$, and $\mathcal{I}_{\sigma^{(2)}_c}^{\sigma^{(1)}_c}$ is the spatial interpolation operator that maps the grid of $\mathcal{G}_2$ to the grid of $\mathcal{G}_1$ (identity mapping if identical grid is used). To formulate the construction of the solution in the initial serial sweep in the proposed solver, let ${\rm HODMD}(A, d, q, t)$ denote the high-order dynamic mode decomposition applied to the snapshot matrix $A=[x(t_0), x(t_1), \cdots,x(t_k)]$ used to predict the data $x(t)$ at time $t$ for $t>t_k$, where $d$ is the order of the high-order dynamic mode decomposition, and $t_i-t_{i-1}=q\tau, ~ i = 1,2,\cdots,k$ where $\tau$ denotes the time step size (which we assume to be uniform for simplicity) used in a time marching scheme. Let $N_{q2}$ denote the number of coarse time steps computed by $\mathcal{G}_2$ on each subinterval. Using Equation \eqref{eq:core1} and \eqref{eq:core2} calculated on two cores in parallel, we compute ${X}_n^0$ as follows.
\begin{equation}\label{eq:init_soln}
{X}_j^0 = \begin{cases}
\mathcal{G}_1(T_j,T_{j-1},X_{j-1}^0), & \text{if}~j=1,2,\cdots,K_t,\\
\mathcal{I}_{\sigma^{(2)}_c}^{\sigma^{(1)}_c}X^0_{j\cdot N_{q2},2}+{\rm HODMD}(\Phi_1-\Phi_2,d_1,q_1,T_j), & \text{if}~j=K_t+1,\cdots,\nc.
\end{cases}
\end{equation}

\subsection{Serial correction sweep}
In the serial correction step, we use HODMD to estimate the correction $\mathcal{G}\left(T_{n},T_{n-1},X^{k}_{n-1}\right)-\mathcal{G}\left(T_{n},T_{n-1},X^{k-1}_{n-1}\right)$ in Equation \eqref{eq:serial_correct}. We use four cores during the first Parareal iteration, i.e. $k=1$ and two cores for $k>1$. For $k=1$ and each $n = k+1, k+2,\cdots,\nc$, calculate Equation \eqref{eq:c1}-\eqref{eq:c4} on four cores in parallel. For $k>1$ and each $n = k+1, k+2,\cdots,\nc$, calculate Equation \eqref{eq:c2} and Equation \eqref{eq:c4} on two cores in parallel.
\begin{align}
{\rm Core}~\#~0: ~ & \mathcal{G}_1(T_{n-1}+l\tau_c^{(2)}, T_{n-1}, X_{n-1}^{k-1}),~{\rm if}~n\geq K_t+1,~k=1~\label{eq:c1}\\
{\rm Core}~\#~1: ~ & \mathcal{G}_1(T_{n-1}+l\tau_c^{(2)}, T_{n-1}, X_{n-1}^k), ~k=1,2,3,\cdots\label{eq:c2}\\
{\rm Core}~\#~2: ~ & \mathcal{G}_2(T_{n},T_{n-1},X_{n-1}^{k-1}), ~{\rm if}~k=1\label{eq:c3}\\
{\rm Core}~\#~3: ~ & \mathcal{G}_2(T_{n},T_{n-1},X_{n-1}^{k}), ~ k=1,2,3,\cdots\label{eq:c4}
\end{align}
Here, $l<N_{q2}$ is a smaller number of time steps than $N_{q2}$ (note $N_{q2}\cdot\tau_c^{(2)}=\Delta T$). In the computation, store the solutions $\mathcal{G}_i(T_{n-1}+j\tau_c^{(2)}, T_{n-1}, X_{n-1}^{k-1})$ and $\mathcal{G}_i(T_{n-1}+j\tau_c^{(2)}, T_{n-1}, X_{n-1}^{k})$ for $j=1,2,\cdots, l$ and $i=1,2$ to form snapshot matrices. Note that in the classic Parareal method, for $k=1$, $\mathcal{G}\left(T_{n},T_{n-1},X^{k-1}_{n-1}\right)$ has been calculated in the initial serial sweep. However, in the proposed solver, since $\mathcal{G}_1(T_j, T_{j-1}, X_{j-1}^0)$ is computed only for $j\leq K_t$, and $\mathcal{G}_2$ is used to obtain $X_{j,2}^0 $ in Equation \eqref{eq:init_soln}, Equation \eqref{eq:c1} and \eqref{eq:c3} are needed for $k=1$, the first Parareal iteration. We remark that for $k>1$, only the calculations given by Equation \eqref{eq:c2} and \eqref{eq:c4} on core \#1 and core \#3 are needed. 

As with the initial serial sweep, we apply $\mathcal{G}_1$ to solve for the equation on a subset of each time interval and apply $\mathcal{G}_2$ for the entire time interval. Once Equation \eqref{eq:c1}-\eqref{eq:c4} are computed, we calculate the difference between the solutions given by the two different solvers on the above subset of the time interval and use its HODMD prediction to estimate the correction term $\mathcal{G}\left(T_{n},T_{n-1},X^{k}_{n-1}\right)-\mathcal{G}\left(T_{n},T_{n-1},X^{k-1}_{n-1}\right)$. Based on the computations performed in Equations \eqref{eq:c1}-\eqref{eq:c4}, form the four following snapshot matrices:
\begin{align}
U_1&= \begin{bmatrix}
X_{n-1}^{k-1},  \mathcal{G}_1(T_{n-1}+\tau_c^{(2)},T_{n-1},X_{n-1}^{k-1}), \cdots, \mathcal{G}_1(T_{n-1}+l\tau_c^{(2)},T_{n-1},X_{n-1}^{k-1})
\end{bmatrix},\\
U_2 &= \begin{bmatrix}
X_{n-1}^k, \mathcal{G}_1(T_{n-1}+\tau_c^{(2)},T_{n-1},X_{n-1}^k), \cdots, \mathcal{G}_1(T_{n-1}+l\tau_c^{(2)},T_{n-1},X_{n-1}^k)
\end{bmatrix},\\
V_1 &=\begin{bmatrix} 
X_{n-1}^{k-1},  \mathcal{I}_{\sigma^{(2)}_c}^{\sigma^{(1)}_c}\mathcal{G}_2(T_{n-1}+\tau_c^{(2)},T_{n-1},X_{n-1}^{k-1}), \cdots, \mathcal{I}_{\sigma^{(2)}_c}^{\sigma^{(1)}_c}\mathcal{G}_2(T_{n-1}+l\tau_c^{(2)},T_{n-1},X_{n-1}^{k-1})
\end{bmatrix},\\
V_2 &=\begin{bmatrix} 
X_{n-1}^{k},  \mathcal{I}_{\sigma^{(2)}_c}^{\sigma^{(1)}_c}\mathcal{G}_2(T_{n-1}+\tau_c^{(2)},T_{n-1},X_{n-1}^{k}), \cdots, \mathcal{I}_{\sigma^{(2)}_c}^{\sigma^{(1)}_c}\mathcal{G}_2(T_{n-1}+l\tau_c^{(2)},T_{n-1},X_{n-1}^{k})
\end{bmatrix}.
\end{align}
The snapshot matrix $U_2-U_1$ stores the correction $\mathcal{G}\left(T_{n-1}+j\tau_c^{(2)},T_{n-1},X^{k}_{n-1}\right)-\mathcal{G}\left(T_{n-1}+j\tau_c^{(2)},T_{n-1},X^{k-1}_{n-1}\right)$ for $\mathcal{G}=\mathcal{G}_1$, $j=1,2,\cdots, l$. The snapshot matrix $V_2-V_1$ stores the numerical approximation of the same quantities, but using a different solver, i.e. $\mathcal{G}=\mathcal{G}_2$. We apply HODMD to their difference $U_2-U_1-(V_2-V_1)$, and add the HODMD prediction of the difference at each $t=T_j$ as a correction to $\mathcal{I}_{\sigma^{(2)}_c}^{\sigma^{(1)}_c}\mathcal{G}_2(T_n,T_{n-1},X_{n-1}^k)-\mathcal{I}_{\sigma^{(2)}_c}^{\sigma^{(1)}_c}\mathcal{G}_2(T_n,T_{n-1},X_{n-1}^{k-1})$, the correction computed using $\mathcal{G}_2$ alone. To formulate the above calculations, let ${X_n^k}^{\prime}=\mathcal{F}\left(T_{n},T_{n-1},X^{k-1}_{n-1}\right)$ be the solution  given by the fine solver calculated in parallel as in Equation \eqref{eq:intermediate}. In the serial correction step of the proposed algorithm, we update $X_n^k$, $U_1$ and $U_2$ by
\begin{equation}
\begin{split}
X_n^k &= X_n^{k'} + \mathcal{I}_{\sigma^{(2)}_c}^{\sigma^{(1)}_c}\mathcal{G}_2(T_n,T_{n-1},X_{n-1}^k)-\mathcal{I}_{\sigma^{(2)}_c}^{\sigma^{(1)}_c}\mathcal{G}_2(T_n,T_{n-1},X_{n-1}^{k-1})\\
&+{\rm HODMD}(U_2-U_1-(V_2-V_1),d_2,q_2,T_n),\\
U_1&\gets U_2,\quad V_1\gets V_2.
\end{split}
\end{equation}
In particular, we note that $l$ may be varied based on the iteration number $k$. In the numerical experiments, we choose $l$ to be increasing as the iteration number $k$ increases. We use the notation $l_k$ to denote the value $l$ used at the $k$th iteration in the outline of the algorithm below in Algorithm \ref{alg:par-dmd} in Section \ref{ssec:alg}.

\subsection{Algorithm}\label{ssec:alg}
We present an outline of the proposed Parareal-HODMD algorithm in Algorithm \ref{alg:par-dmd}.
\begin{breakablealgorithm}
\caption{The Parareal-HODMD algorithm}\label{alg:par-dmd}
\begin{algorithmic}[1]

\INPUT 

$x_0$, which is the initial condition of the IVP of the form Eq. \eqref{eq:ode}
\vspace{.1in}

Number of cores $\nc$ 
\vspace{.1in}

Parameters of the fine solver: $\tau_f, \sigma_f.$

Parameters of the coarse solvers: $K_t<\nc, \tau_c^{(1)}, \tau_c^{(2)}, N_{q1}, N_{q2}, \sigma_c^{(1)}, \sigma_c^{(2)}, r_q, T_j=jT/\nc, d_1,d_2,q_1,q_2, \{l_k\}_k ~ {\rm where}~l_k<N_{q2}~{\rm and}~\{l_k\}_k~\text{forms an increasing sequence}$

\vspace{.1in}

\OUTPUT $x(T_n)$, the estimated solution to Eq. \eqref{eq:ode} at time $T_n=n\cdot T/\nc$
\vspace{.1in}

\hspace{-1.5cm} \fcolorbox{lightgray}{lightgray}{\texttt{\# the initial serial sweep}}
\vspace{.05in}

\hspace{-1.5cm} Initialize $X_0^0\gets x_0$
\vspace{.05in}

\hspace{-1.5cm} Perform the following two calculations on two cores \textbf{in parallel}:

\hspace{-1.5cm} Core \#0:

\For{$j=~1,~2\cdots,~K_t$} 
\State Calculate $\mathcal{G}_1(T_j, T_{j-1}, X_{j-1}^0)$
\EndFor

\State Form $\Phi_1 = \begin{bmatrix} X^0_{0,1}, X^0_{1,1}, \cdots, X^0_{p,1}\end{bmatrix},~p=N_{q1}\cdot K_t$

\hspace{-1.5cm} Core \#1:

\For{$j=~1,~2\cdots,~\nc$} 
\State Calculate $\mathcal{G}_2(T_j, T_{j-1}, X_{j-1}^0)$ 
\EndFor

\State $\Phi_2 = \begin{bmatrix} \mathcal{I}_{\sigma^{(2)}_c}^{\sigma^{(1)}_c}X^0_{0,2}, \mathcal{I}_{\sigma^{(2)}_c}^{\sigma^{(1)}_c}X^0_{1,2}, \cdots, \mathcal{I}_{\sigma^{(2)}_c}^{\sigma^{(1)}_c}X^0_{p,2}\end{bmatrix}$

\If {$1\leq j \leq K_t$}
\State{$\vc X_j^0\gets\mathcal{G}_1(T_j,T_{j-1},X_{j-1}^0)$}
\Else 
\State{$\vc X_j^0\gets\mathcal{I}_{\sigma^{(2)}_c}^{\sigma^{(1)}_c}X^0_{j\cdot N_{q2},2}+{\rm HODMD}(\Phi_1-\Phi_2,d_1,q_1,T_j)$}
\EndIf
 
\hspace{-1.5cm}  \fcolorbox{lightgray}{lightgray}{\texttt{\# the main for-loop}}
 \For{$k=1,~2,~3, ~\cdots$} \quad 
 
 \State Calculate $\left\{{\vc X_n^k}^{\prime}=\mathcal{F}\left(T_{n},T_{n-1},\vc X_{n-1}^{k-1}\right)\right\}_{n=k}^{\nc}$ \textbf{in parallel} \quad \fcolorbox{lightgray}{lightgray}{\texttt{\# parallel sweep}}
   
\If {$k>1$}
  
\State $\vc X_n^k\gets\vc X_n^{k-1}$ for $n=1,~2,\cdots,~k-1$
  
\EndIf

\State $\vc X_k^k \gets {\vc X_k^k}^{\prime}$

\fcolorbox{lightgray}{lightgray}{\texttt{\# serial correction}}

Perform the following computations on four cores \textbf{in parallel}:

\For{$n=k+1,~k+2,\cdots,~\nc$} 

Core \#0:
\If {$k=1~{\rm and}~n\geq K_t+1$}
\State Calculate $\mathcal{G}_1(T_{n-1}+l_k\tau_c^{(2)}, T_{n-1}, X_{n-1}^{k-1})$
\State Form $U_1= \begin{bmatrix}
X_{n-1}^{k-1},  \mathcal{G}_1(T_{n-1}+\tau_c^{(2)},T_{n-1},X_{n-1}^{k-1}), \cdots, \mathcal{G}_1(T_{n-1}+l_k\tau_c^{(2)},T_{n-1},X_{n-1}^{k-1})
\end{bmatrix}$
\EndIf

Core \#1:
\State Calculate $\mathcal{G}_1(T_{n-1}+l_k\tau_c^{(2)}, T_{n-1}, X_{n-1}^{k})$

\State Form $U_2 = \begin{bmatrix}
X_{n-1}^k, \mathcal{G}_1(T_{n-1}+\tau_c^{(2)},T_{n-1},X_{n-1}^k), \cdots, \mathcal{G}_1(T_{n-1}+l_k\tau_c^{(2)},T_{n-1},X_{n-1}^k)
\end{bmatrix}$

Core \#2:
\If {$k=1$}
\State Calculate $\mathcal{I}_{\sigma^{(2)}_c}^{\sigma^{(1)}_c}\mathcal{G}_2(T_{n-1}+N_{q2}\tau_c^{(2)},T_{n-1},X_{n-1}^{k-1})$
\State Form $V_1=\begin{bmatrix} 
X_{n-1}^{k-1},  \mathcal{I}_{\sigma^{(2)}_c}^{\sigma^{(1)}_c}\mathcal{G}_2(T_{n-1}+\tau_c^{(2)},T_{n-1},X_{n-1}^{k-1}), \cdots, \mathcal{I}_{\sigma^{(2)}_c}^{\sigma^{(1)}_c}\mathcal{G}_2(T_{n-1}+l_k\tau_c^{(2)},T_{n-1},X_{n-1}^{k-1})
\end{bmatrix}$
\EndIf

Core \#3:
\State Calculate $\mathcal{I}_{\sigma^{(2)}_c}^{\sigma^{(1)}_c}\mathcal{G}_2(T_{n},T_{n-1},X_{n-1}^{k})$

\State Form $V_2=\begin{bmatrix} 
X_{n-1}^{k},  \mathcal{I}_{\sigma^{(2)}_c}^{\sigma^{(1)}_c}\mathcal{G}_2(T_{n-1}+\tau_c^{(2)},T_{n-1},X_{n-1}^{k}), \cdots, \mathcal{I}_{\sigma^{(2)}_c}^{\sigma^{(1)}_c}\mathcal{G}_2(T_{n-1}+l_k\tau_c^{(2)},T_{n-1},X_{n-1}^{k})
\end{bmatrix}$

Core \#0:
\State Update
\begin{equation*}
\begin{split}
X_n^k \gets & X_n^{k'} + \mathcal{I}_{\sigma^2_c}^{\sigma^1_c}\mathcal{G}_2(T_n,T_{n-1},X_{n-1}^k)-\mathcal{I}_{\sigma^2_c}^{\sigma^1_c}\mathcal{G}_2(T_n,T_{n-1},X_{n-1}^{k-1})\\
&+{\rm HODMD}(U_2-U_1-(V_2-V_1),d_2,q_2,T_n)
\end{split}
\end{equation*}
\State $U_1\gets U_2$, $V_1\gets V_2$
\EndFor
\EndFor
 
\end{algorithmic}
\end{breakablealgorithm}

\section{Numerical examples}\label{sec:examples}
In this section, we present numerical results of Algorithm \ref{alg:par-dmd}, applied to different fluid simulations, including the fluid-structure interaction problems of microswimmers of different configurations and the dynamics of thin liquid films on solid substrates. Concretely, we test the algorithm on two scenarios, one with Lagrangian description of the fluid problem solved using an explicit time marching scheme which has a fixed spatial complexity at each time step, the other with Eulerian description of the fluid problem solved using an implicit time marching scheme, which is solved with an iterative method. In Section \ref{ssec:fsi}, we present the first scenario, in which we apply the algorithm to two examples of bio-inspired fluid simulations. We describe the model of microswimmers immersed in incompressible Newtonian fluid at zero Reynolds number and the Method of Regularized Stokeslets (MRS). Specifically, we consider microswimmers of two structures, i.e. a rod-like swimmer and a spherical swimmer. In \ref{sssec:rod-swimmer}, we apply Algorithm \ref{alg:par-dmd} to simulate the motion of a slender, elastic rod, which is often used to model biological structures such as tails of sperm and bacterial flagella. In \ref{sssec:sphere}, we simulate the dynamics of an elastic sphere, which can be used to model a vesicle placed in a shear flow. The two examples are characterized by relatively long-time simulations and small numbers of spatial grid points. Both examples are solved using an explicit time scheme in the fine and the coarse solver of the algorithms. In Section \ref{ssec:film}, we present the second scenario, where we use Algorithm \ref{alg:par-dmd} to solve for the evolution of thin liquid film described by lubrication equation on a chemically heterogeneous two-dimensional substrate. The example is solved using an implicit time marching scheme and an iterative approach for the discretized equation, and is characterized by a larger number of spatial grid points and spatial complexity compared to the examples in Section \ref{ssec:fsi}.

In all of the numerical experiments, we apply both the original Parareal algorithm described in Section \ref{sec:parareal} and the proposed Parareal-HODMD algorithm in Section \ref{sec:parareal-hodmd}. To examine the accuracy and convergence of both algorithms, we compare the solution given by the two Parareal methods to the solution given by the fine solver $\mathcal{F}$ run in serial. In all of the experiments, we solve a differential equation on a given time domain $t\in(0,T]$. Let $\vc x_{i}^k$ and $\mathbf{x}_{i}^{\mathcal{F}}$ denote the numerical solution of the differential equation at final time $t=T$, i.e. $\mathbf{x}_i(T)$ calculated by a Parareal algorithm and the fine solver $\mathcal{F}$ run in serial respectively. In the experiments, we examine two metrices, the true relative error and relative increment obtained at the $k$th iteration of Algorithm \ref{alg:par-dmd}:
\begin{linenomath*}
\begin{equation}\label{eq:true_err}
\eta^k=
\max_{i=1,~2,~\cdots,~N_{\sigma}}\frac{\left\|\mathbf{x}_i^k-\mathbf{x}_i^{\mathcal{F}}\right\|_2}{\left\|\mathbf{x}_i^{\mathcal{F}}\right\|_2}~~\text{and}~~
\widetilde{\eta}^k=\max_{i=1,~2,~\cdots,~N_{\sigma}}\frac{\left\|\mathbf{x}^k_i-\mathbf{x}^{k-1}_i\right\|_2}{\left\|\mathbf{x}^k_i\right\|_2}~~\text{for}~~k=1,~2,~\cdots
\end{equation}
\end{linenomath*}
where $N_{\sigma}$ is the number of grid points. In practice, the serial solution is not known, the relative increment which measures the difference between consecutive iterates of Algorithm \ref{alg:par-dmd} is used in the stopping criterion of Algorithm \ref{alg:par-dmd}. All the computational results are obtained using mpi4py in Python on Intel\textsuperscript{\textregistered} Xeon\textsuperscript{\textregistered} CPU E5-262 v3. 

\subsection{Microswimmers in Stokes flow}\label{ssec:fsi}
The motion of microswimmers at the scale of zero Reynolds number immersed in incompressible Newtonian fluid can be described by Stokes equation:
\begin{linenomath*}
\begin{subequations}
\begin{alignat}{2}
\vc 0&=-\nabla p+\mu\Delta \vc u+\mathbf{f}^{\text{s}}\label{eq:stokes}\\
0&=\nabla\cdot\vc u,\label{eq:incompress}
\end{alignat}
\end{subequations}
\end{linenomath*}
where $\mu$ is the fluid's viscosity, $\vc u = \vc u(\vc x)\in\mathbb{R}^3$ and $p = p(\vc x)\in\mathbb{R}$ are the velocity and pressure at $\vc x\in\mathbb{R}^3$, and $\mathbf{f}^{\text{s}}\in\mathbb{R}^3$ is an applied body force. The fundamental solution of Equations (\ref{eq:stokes}) and (\ref{eq:incompress}), known as a Stokeslet, gives the velocity solution for a single point force $\vc f\in\mathbb{R}^3$ at $\vc y$, $\mathbf{f}^{\text{s}}=\vc f\delta(r)$, where $\delta$ denotes the Dirac Delta function, and $r$ denotes the radial distance $|\vc x-\vc y|$ between $\vc x$ and $\vc y$. Evaluating the velocities where the force is exerted is numerically challenging due to the singularity at the point. To solve this problem, Method of Regularized Stokeslets (MRS) is employed \cite{Cortez2001,CortezEtAl2005}, where the singularity is removed by replacing the Dirac Delta distribution $\delta$ by a radially symmetric, smooth function called a blob function. An example of a frequently used family of blob functions is given by 
\begin{linenomath*}
\begin{equation}\label{eq:blob}
\phi_{\eps}(r)=\frac{15\eps^4}{8\pi(r^2+\eps^2)^{7/2}},
\end{equation}
\end{linenomath*}
where $\eps>0$ is a parameter describing the width of the blob. In physics and biology applications, $\eps$ is often chosen to be the physical radius or thickness of the structure modeled. As $\eps$ approaches zero, $\phi_{\eps}$ converges to $\delta$ in the distribution sense. For $\vc{f}_s=\vc{f}\phi_{\eps}(r)$, the solution to Equation (\ref{eq:stokes})-(\ref{eq:incompress}) is given by
\begin{linenomath*}
\begin{equation}\label{eq:uregstoke}
\vc u(\vc x) = \frac{1}{\mu}\left(\vphantom{\frac{1}{2}}H_1(r)\vc{f}+H_2(r)\left(\vc{f}\cdot (\vc{x}-\vc{y})\right)(\vc{x}-\vc{y})\right),
\end{equation}
\end{linenomath*}
where $H_1$ and $H_2$ are radially symmetric, smooth functions whose expressions depend on the choice of $\phi_{\eps}$. Eq. (\ref{eq:uregstoke}) is referred to as a regularized Stokeslet, and can be evaluated for $\vc x\in\mathbb{R}^3$. When there are ${N_{\sigma}}$ grid points where forces are exerted, the velocity $\vc{u}$ at any $\vc{x}$ can be expressed by the sum of the effects induced by the ${N_{\sigma}}$ point forces at $\{\vc x_{i,\ell-1}\}_{i=1}^{N_{\sigma}}$ as follows:
\begin{linenomath*}
\begin{equation}\label{eq:usumregstoke}
\vc u(\vc x) =\sum_{i=1}^{N_{\sigma}} \frac{1}{\mu}\left(\vphantom{\frac{1}{2}}H_1(r_i)\vc{f}_i+H_2(r_i)\left(\vc{f}_i\cdot (\vc{x}-\vc{x}_i)\right)(\vc{x}-\vc{x}_i)\right),
\end{equation}
\end{linenomath*}
for $\vc x\in\mathbb{R}^3$. For a fluid bounded by an infinite, planar, and stationary wall, the regularized solution to Equation (\ref{eq:stokes})-(\ref{eq:incompress}) has been derived in \cite{AinleyEtAl2008} using the method of images. By Equation (\ref{eq:usumregstoke}), the complexity of computing the velocity at $\vc u(\vc x_i)$ for $i=1,2,\cdots,N_{\sigma}$ at a time is $O\left(N_{\sigma}^2\right)$. If an explicit time marching scheme is applied to compute the position of $N_{\sigma}$ points for $N_{\tau}$ time steps, the computational cost is $O(N_{\sigma}^2N_{\tau})$. 

\subsubsection{Single rod-like swimmer}\label{sssec:rod-swimmer}
We consider a rod-like swimmer modeled by a Kirchhoff rod, which has been used in \cite{OlsonEtAl2013,Lim2010,Olson2014,HuangEtAl2018,CarichinoOlson2019} for the computational modeling of flagellar swimmer. We describe the rod using its centerline which is a space curve and assume its flagellar waveform as sinusoidal based on previous experimental studies \cite{Ho2001HyperactivationOM,Smith2009BendPI}. We prescribe the rod to propagate a planar sinusoidal wave by prescribing it with a preferred strain twist vector in \cite{OlsonEtAl2013}, given by
 \begin{equation}
\left(\Omega_1,\Omega_2,\Omega_3\right)=\left(0,-k^2A\sin(ks+f t),0\right),
\end{equation}
where $s$ is the arclength, $A$ is the amplitude, $f$ is the frequency, and $k=2\pi/\lambda$ for the wavelength $\lambda$. We assume the rod is in the semi-infinite fluid domain $\{(x,y,z)\in\mathbb{R}^3\bigr | z\ge 0\}$ bounded by an infinite, planar, and stationary wall located at $z=0$, on which the flow vanishes. In addition, at time $t=0$, the rods are initialized to be straight in a plane parallel to the $z$-plane with a distance $d$ above the wall at $z=0$. The parameter values that describe the material properties of the rod are chosen based on the ranges of the tails of human sperm considered in \cite{CarichinoOlson2019}. The parameter values used in this experiment are given in Table \ref{tb:wave}. 

\begin{table}[htbp!]
\centering
\caption{Parameters of the material properties and the waveform of the rod}
\begin{tabular}{r r r}
\hline
Rod length, $L$ ($\mu m$) & & 60 \\
Amplitude, $A$ ($\mu m$) & & 1 \\
Frequency, $f$ (Hz) && $40\pi$\\
Wavelength, $\lambda$ ($\mu m$)  && 30\\
Bending modulus, $a_1=a_2$ ($g\cdot\mu m^3\cdot s^{-2}$)  && $1$\\
Twist modulus, $a_3$ ($g\cdot\mu m^3\cdot s^{-2}$) & & $1$ \\
Shear modulus, $b_1=b_2$ ($g\cdot\mu m\cdot s^{-2}$) & & $0.6$ \\
Stretch modulus, $b_3$ ($g\cdot\mu m\cdot s^{-2}$) & & $0.6$ \\
Distance to wall, $d$ ($\mu m$)  & & $10$\\
Fluid viscosity, $\mu$ ($g\cdot\mu m^{-1}\cdot s^{-1}$)  & & $10^{-6}$\\
\hline
\end{tabular}
\label{tb:wave}
\end{table}

\begin{figure}[htbp!]
\centering
\includegraphics[height=1.3in]{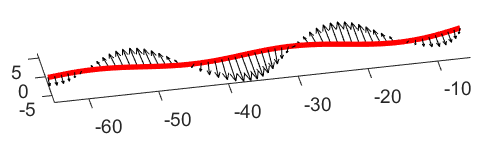}
\caption{Configuration of the rod swimmer and the velocity along the rod at $t=1$}
\label{fig:rod_visualize}
\end{figure}

We apply both the original Parareal algorithm and the proposed Parareal-HODMD algorithm to the simulation of a single rod swimmer in Stokes flow, initialized as a straight rod with distance $d=10$ to the wall at $z=0$. For both algorithms, we uniformly discretize the rod using $N_{\sigma}=301$ points along the length of the rod, with $\sigma_f=L/(N_{\sigma}-1)=0.2\mu$ m. The explicit two-stage Runge-Kutta method (RK2) is used to march forward in time with a step size $\tau_f=10^{-6}$ s. Figure \ref{fig:rod_visualize} shows the rod and the velocity along the rod with parameters given in Table \ref{tb:wave} solved for $t=1$, using RK2 with regularization parameter $\epsilon=5\sigma_f$.

First, we show the convergence of the proposed algorithm. We consider the case of a single rod with two different values of radius, corresponding to two different regularization parameters $\eps$. We perform two experiments where we set the regularization parameter of MRS to be $\epsilon=5\sigma_f$ and $\epsilon=3\sigma_f$. A smaller regularization parameter leads to a stiffer problem, which has a more stringent requirement on the size of the time step. In both experiments, we solve for the motion of the rod on time interval $[0,T]$ for $T=0.1$ with $\nc=50$ cores. The coarse solver is chosen to be the forward Euler's method in both algorithms. Here, we summarize the parameters used in $\mathcal{G}_1$ and $\mathcal{G}_2$ of the Parareal-HODMD algorithm for both cases. For $\epsilon=5\sigma_f$, we set $\tau_c^{(1)}=8\tau_f$, and $\tau_c^{(2)}=2\tau_c^{(1)}$ with $N_q=250$ and $r_q=2$. For $\epsilon=3\sigma_f$, the case of a smaller regularization parameter, the time step required by the explicit time marching scheme is extremely restrictive. For $\epsilon=3\sigma_f$, we use $\tau_c^{(1)}=\tau_f$ and $\tau_c^{(2)}=2\tau_f$, which gives $N_q=2000$ and $r_q=2$. For both regularization parameters, we run the algorithm for three iterations, with $(l_1,l_2,l_3)$ specified in Table \ref{tb:par-param}. We note that due to the coarser grid used in both space and time dimension in $\mathcal{G}_2$, the theoretical computational cost of the coarse solver $\mathcal{G}_2$ is less than the cost of $\mathcal{G}_1$, which implies the cost of the serial calculation is determined by the cost of $\mathcal{G}_1$. 

\begin{table}[htbp!]
\centering
\caption{Parameters used in the two coarse solvers of the Parareal-HODMD algorithm}
\begin{tabular}{c c c}
Parameter & $\eps=5\sigma_f$ & $\eps=3\sigma_f$\\
\hline
$\tau_c^{(1)}$ & $8\cdot 10^{-6}$ & $10^{-6}$ \\
$\tau_c^{(2)}$ & $16\cdot 10^{-6}$ & $2\cdot10^{-6}$ \\
$\sigma_c^{(1)}$ & $0.2$ & $0.2$\\
$\sigma_c^{(2)}$ & $1.2$ & $1.2$\\
$K_t$ & 30 & 30\\
$d_1$ & 10 & 10\\
$d_2$ & 12 & 12\\
$q_1$ & 25 & 50\\
$q_2$ & 1 & 2\\
$(l_1,l_2,l_3)$ & (110,130,240) & (880,1560,1920)\\
\hline
\end{tabular}
\label{tb:par-param}
\end{table}

\begin{figure}[htbp!]
\centering
 \begin{subfigure}[b]{0.45\linewidth}
 \centering
 \includegraphics[height=2in]{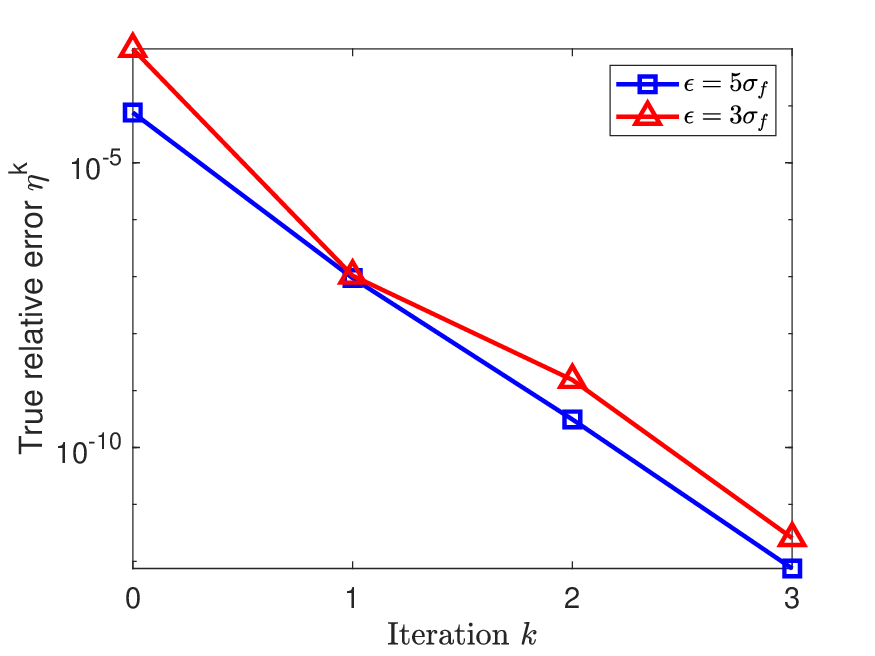}
\caption{}\label{fig:true_rel_err}
\end{subfigure}
 \begin{subfigure}[b]{0.45\linewidth}
 \centering
\includegraphics[height=2in]{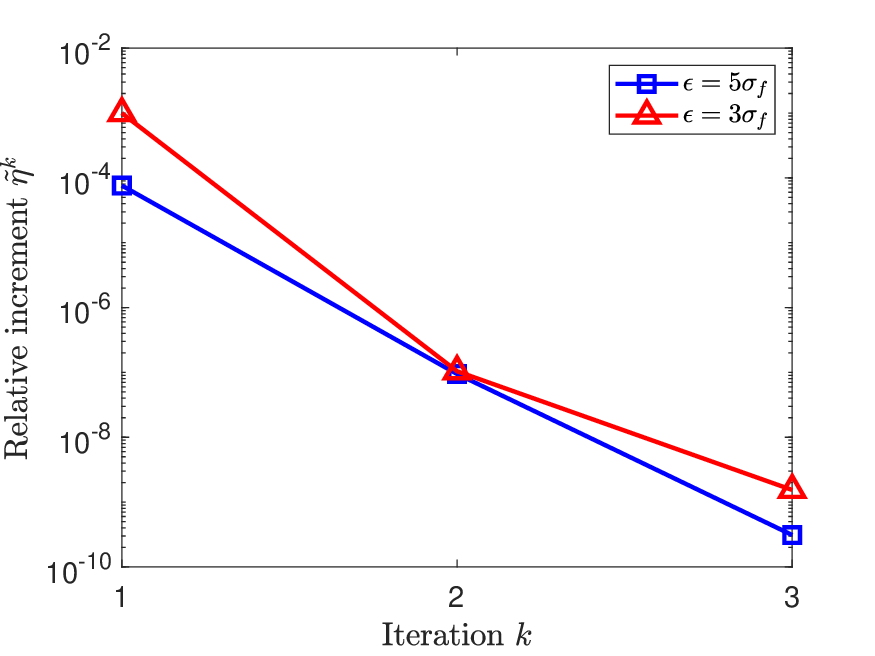}
\caption{}\label{fig:rel_increm}
\end{subfigure}
\caption{(a) Decay of true relative error $\eta^k$ as the number of iteration increases (b) Decay of relative increment $\tilde{\eta}^k$ as the number of iteration increases}
\label{fig:results1}
\end{figure}

In Figure \ref{fig:results1} (a) and (b), we show the decay of true relative error $\eta^k$ and relative increment $\tilde{\eta}^k$ (in semi-log scale) obtained in the numerical experiments for $\eps=5\sigma_f$ and $\eps=3\sigma_f$, where we use $\eta^0$ to denote the true relative error of the solution after the initial serial sweep. From the numerical results in Figure \ref{fig:results1} (a) and (b), we observe a consistent decay of both true relative error and relative increment as the number of iterations increases. The true relative error $\eta^k$ falls below $10^{-6}$ immediately after the first iteration. In particular, we remark that due to the stability requirement of the problem, with the same fine solver, if we simply apply the classic Parareal method using a coarse solver with a spatial grid spacing $\sigma_c=6\sigma_f$ and $\tau_c=\tau_f$, the algorithm does not converge. However, with the application of Parareal-HODMD algorithm, we are able to achieve the convergence of the solution with the incorporation of both $\mathcal{G}_1$ and $\mathcal{G}_2$, where $\sigma_c=6\sigma_f$ and $\tau_c=\tau_f$ are used in $\mathcal{G}_2$.

Next, to demonstrate the time cost and accuracy of the Parareal-HODMD algorithm, we apply the Parareal-HODMD algorithm to the simulation of a rod swimmer using three sets of different parameters, and compare its performance with the original Parareal algorithm, for $t\in[0,T]$ with $T=1$. We consider two cases, $\epsilon=3\sigma_f$ and $\epsilon=5\sigma_f$. In both algorithms, RK2 is used in the fine solver and the forward Euler is used as the numerical scheme in the coarse solver. One iteration is performed, and 50 cores are used for both algorithms. First, we present the total runtime and $\eta^1$ of the two algorithms for the experiment with $\epsilon=3\sigma_f$ in Table \ref{tab:runtime}. In the original Parareal algorithm, we set $\sigma_c=\sigma_f$ and $\tau_c=\tau_f$, as with the parameters used in the experiment with $T=0.1$ described above. In the Parareal-HODMD algorithm, we demonstrate the flexibility and robustness of the algorithm by performing three numerical experiments where we adjust the accuracy of the coarse solvers by varying $K_t$ and $l_1$. Specifically, we increase the accuracy of the HODMD prediction in the coarse solver by increasing $K_t$ and $l_1$. For the three sets of parameters used in the Parareal-HODMD algorithm, we fix $\tau_c^{(1)}=\tau_f$, $\tau_c^{(2)}=2\tau_f$, $\sigma_c^{(1)}=\sigma_f$, and $\sigma_c^{(2)}=6\sigma_f$. Note that for a fair comparison, the $\mathcal{G}_1$ solver of the Parareal-HODMD algorithm uses the same numerical scheme and grid resolution as the coarse solver of the original Parareal algorithm. 

First, we show the results obtained in the stiffer case, $\epsilon=3\sigma_f$. Since the coarse solver based on an explicit time marching scheme allows for little relaxation on the coarsening of time step size, as can be seen in Table \ref{tab:runtime}, the original Parareal algorithm based on such coarse solver has an extremely long runtime and low parallel speedup due to the high cost of the serial calculation involved, regardless of the stopping criterion and desired error tolerance. While the algorithm yields a highly accurate solution with $\eta^1<10^{-10}$ after one iteration, the rigid time step restriction of the problem severely limits the efficiency of the algorithm. In contrast, with the incorporation of the HODMD approximation in the Parareal-HODMD algorithm, for a less stringent stopping criterion, we are able to reduce the total runtime and improve the parallel speedup of the algorithm by constructing coarse solvers of various accuracies and time costs unattainable by the original Parareal algorithm. The total runtime of the three simulations calculated by Parareal-HODMD is approximately 56\%, 62\% and 71\% of the runtime of the original Parareal algorithm, with $\eta^{1}$ of order $10^{-5}$, $10^{-7}$ and $10^{-8}$ respectively. For a desired accuracy $\eta^k<10^{-4}$ and $\eta^k<10^{-6}$, the Parareal-HODMD algorithm achieves a speedup of approximately 5.37 and 4.79 respectively, compared to a mere speedup of 2.98 of the original Parareal algorithm. For a stiff problem like $\epsilon=3\sigma_f$, the original Parareal method based on classic solvers offers little room for coarsening time steps and thus limited room for achieving higher speedups regardless of the choice of stopping criterion. However, the Parareal-HODMD algorithm enables the achieving of higher speedup by the increased flexibility of the construction of the coarse approximation used by the proposed solver.

\begin{table}[htbp!]
\caption{Runtime and accuracy of the Parareal-HODMD algorithm and the original Parareal algorithm for $\epsilon=3\sigma_f$}
\centering
\begin{tabular}{c|c|c|c|c}
\hline
Algorithm & Parameters &  Runtime (s) & Speedup & $\eta^1$ \\
\hline
Original Parareal & $\sigma_c=\sigma_f, \tau_c=\tau_f$ & $63812.05$ & 2.98& $5.91\cdot 10^{-11}$\\
\hline
\multirow{3}{*}{Parareal-HODMD} & $K_t=15$, $l_1=700$ & 35460.88 & 5.37& $1.35\cdot 10^{-5}$\\
\hhline{~----}& $K_t=30$, $l_1=880$ & 39696.92 & 4.79& $1.04\cdot 10^{-7}$\\
\hhline{~----}& $K_t=30$, $l_1=1160$ & 45465.26 & 4.19& $4.90\cdot 10^{-8}$\\
\hline
\end{tabular}
\label{tab:runtime}
\end{table}


Next, in Table \ref{tab:runtime2}, we show the total runtime and $\eta^1$ obtained using the two algorithms for the less stiff case $\epsilon=5\sigma_f$. For $\epsilon=5\sigma_f$, if one employs the original Parareal algorithm, as with the stiffer case $\epsilon=3\sigma_f$ described above, when a coarser time step is used in the Euler's scheme in the coarse solver in the classic Parareal algorithm, the fine spatial grid has to be used in the coarse solver for the algorithm to converge, otherwise the solution does not converge. Hence, the fine spatial grid is used in both the fine and the coarse solver of the classic Parareal algorithm. In the Parareal-HODMD algorithm, we employ a coarser spatial grid in $\mathcal{G}_2$. Specifically, for the three sets of parameters used in the Parareal-HODMD algorithm, we fix $\tau_c^{(1)}=8\tau_f$, $\tau_c^{(2)}=2\tau_c^{(1)}$, $\sigma_c^{(1)}=\sigma_f$ and $\sigma_c^{(2)}=6\sigma_f$. For a fair comparison, the solver $\mathcal{G}_1$ uses the same numerical scheme and grid resolution as the coarse solver of the original Parareal algorithm. We increase the accuracy of the algorithm by increasing $K_t$ and $l_1$ in the three experiments. From Table \ref{tab:runtime2}, since a highly accurate coarse solver is used in the original Parareal algorithm, the algorithm yields a true relative error less than $10^{-9}$ after one iteration with a speedup of 7.27. As we adjust the DMD parameters of the Parareal-HODMD algorithm in the three experiments, we obtain solutions with increasing accuracy after one iteration, with a speedup of 24.67, 19.29 and 17.01 respectively, which is significantly greater than the the speedup of the original Parareal algorithm. In particular, we note that the third set of parameters results in a solution which is approximately just one order of magnitude larger than the solution given by the original Parareal algorithm. However, the speedup achieved by the proposed algorithm is approximately $2.34$ times of that of the original Parareal algorithm.
\begin{table}[htbp!]
\caption{Time cost and accuracy of the Parareal-HODMD algorithm and the original Parareal algorithm $\epsilon=5\sigma_f$}
\centering
\begin{tabular}{c|c|c|c|c}
\hline
Algorithm & Parameters & Runtime (s) & Speedup & $\eta^1$ \\
\hline
Original Parareal & $\sigma_c=\sigma_f, \tau_c=8\tau_f$ & $ 16779.27$ & 7.27 & $8.24\cdot 10^{-10}$\\
\hline
\multirow{3}{*}{Parareal-HODMD} & $K_t=15$, $l_1=40$ & 4943.33 & 24.67 & $1.61\cdot 10^{-3}$\\
\hhline{~----}& $K_t=30$, $l_1=55$ &  6322.90 & 19.29 & $1.01\cdot 10^{-7}$ \\
\hhline{~----}& $K_t=35$, $l_1=70$ & 7169.39 & 17.01 & $9.76\cdot 10^{-9}$ \\
\hline
\end{tabular}
\label{tab:runtime2}
\end{table}

\subsubsection{Elastic sphere in the shear flow}\label{sssec:sphere}
The dynamics of spherical and ellipsoidal structures in fluid has been studied extensively due to its vast applications in engineering and sciences \cite{2011Slow,2020Numerical,Fawcett2001Scattering,2022Soft,Friedman2017Dynamics,2006The}. For example, the motion of spherical and ellipsoidal structures in fluid has been used to model the dynamics of suspension of particles \cite{Yang2016NumericalSO,2006The}, and the deformation and motion of vesicles \cite{Veerapaneni2009ABI,Martin1996Fluid}. In this section, we apply Algorithm \ref{alg:par-dmd} to the simulation of an elastic spherical surface in a simple shear flow in free space. The surface of the sphere is modeled by a network of connected Hookean springs. The background shear flow causes the surface to undergo deformations. In this experiment, we consider a spherical surface $x^2+y^2+z^2=1$ placed in a shear flow with background fluid velocity $\vc u_{\infty}(x,y,z)=(0.1z,0,0)^T$. The mesh of the surface consists of close-to-uniform triangles. Each edge of the triangles is modeled as a Hookean spring with a Spring constant $K$. Similar force models have been considered in \cite{bouzarth2010multirate,jackson2015applications,navot1998elastic} to simulate the motion of spherical surfaces in a viscous shear flow. The fluid velocity at any point $\vc x$ is given by $\vc u_{\infty}(\vc x)+\vc u^s(\vc x)$, where $\vc u^s(\vc x)$ is induced by the spring forces and calculated using Equation \eqref{eq:usumregstoke}. 

\begin{figure}[htbp!]
\centering
 \begin{subfigure}[b]{0.45\linewidth}
 \centering
 \includegraphics[height=1.5in]{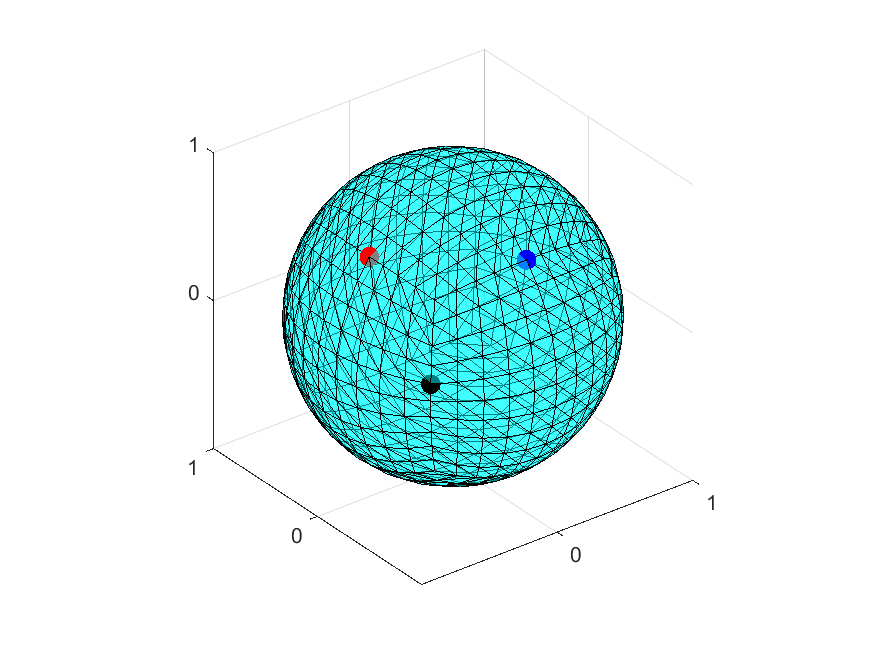}
\caption{}\label{fig:sphere_tracer}
\end{subfigure}
 \begin{subfigure}[b]{0.45\linewidth}
 \centering
\includegraphics[height=1.5in]{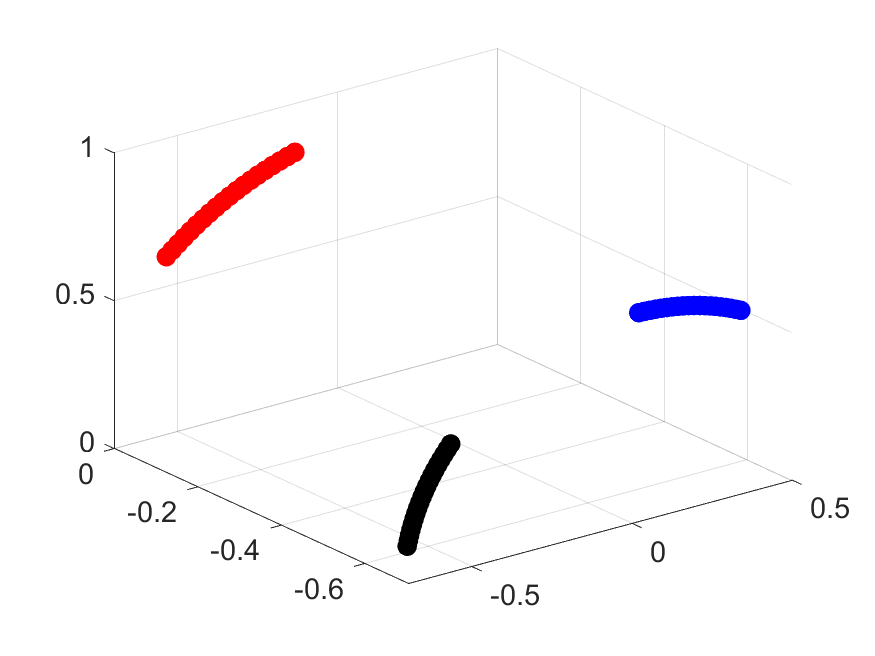}
\caption{}\label{fig:tracer_path}
\end{subfigure}\\
 \begin{subfigure}[b]{0.45\linewidth}
 \centering
\includegraphics[height=2in]{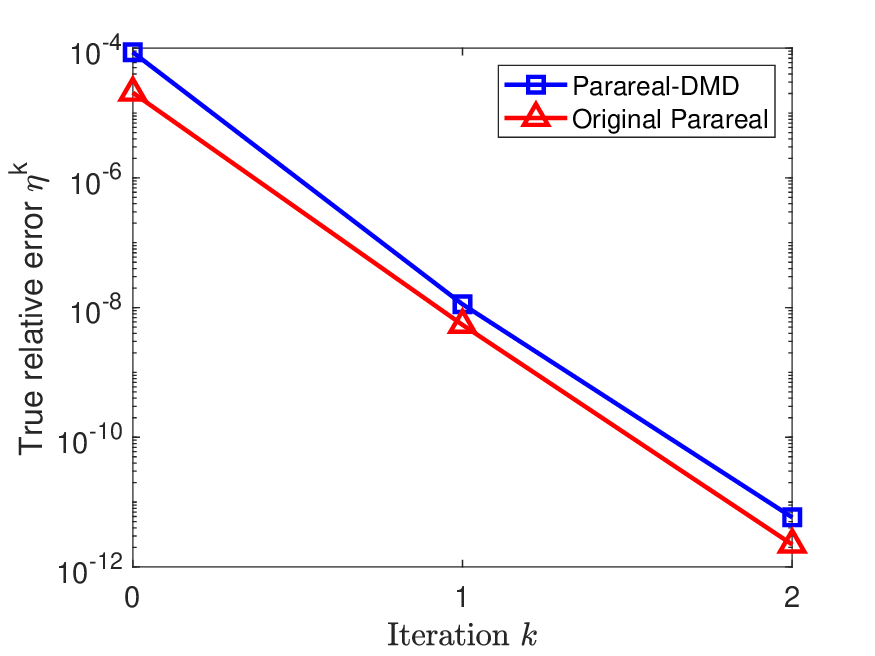}
\caption{}\label{fig:sphere_err}
\end{subfigure}
\caption{(a) Three colored tracers marked on the surface of the sphere (b) The trajectory of the three tracers for $t\in[0, 100]$ (c) True relative error $\eta^k$ of the Parareal-HODMD and the original Parareal algorithm}
\label{fig:visualize_sphere}
\end{figure}


In the numerical experiment, we consider a unit sphere with $N_{\sigma}=602$ points with a triangular mesh consisting of 1200 triangles. Let $h_{\max}$ denote the maximum edge length of the triangle mesh. We use a regularization parameter of $\epsilon=0.3h_{\max}$. We set the spring constant of the Hooke's law to be $K=0.1$, which is a relatively stiff case and the fluid viscosity $\mu=10^{-3}$. For both algorithms, we solve for the motion of the sphere for time $t\in[0, T]$ for $T=10$ and $T=100$ using RK2 as the fine numerical scheme and forward Euler as the coarse numerical scheme. Specifically, in both algorithms, we use a time step of $\tau_f=10^{-3}$ in the fine solver. For the coarse solvers, in the original Parareal method, we set $\tau_c=4\cdot 10^{-3}$; in the Parareal-HODMD method, we set $\tau_c^{(1)}=4\cdot10^{-4}$, and $\tau_c^{(2)}=2\tau_c^{(1)}$, which gives $N_q=50$ for the case $T=10$ and $N_q=500$ for the case $T=100$. We use the same spatial mesh in both the fine solver and coarse solver in both algorithms. We note that in the Parareal-HODMD algorithm, due to the stability requirment of the problem, directly applying the forward Euler's method with $\tau_c^{(2)}=2\tau_c^{(1)}=8\cdot10^{-3}$ is not allowed. In order to construct a coarse solver that uses a time step of such size, we employ a larger regularization parameter $\epsilon=0.5h_{\max}$ in $\mathcal{G}_2$ of the Parareal-HODMD algorithm, which makes it less stiff and at the same time less accurate due to both a coarser temporal resolution as well as a different regularization parameter used. We run the algorithm for one iteration and set $K_t=25$, $d_1=d_2=5$, $l_1=25$, $q_2=1$ for both of the experiments $T=10$ and $T=100$. A value of $q_1=5$ is used for $T=10$, and a value of $q_1=50$ is used for $T=100$. As with the experiment of the rod-like swimmer, we remark that if $\mathcal{G}_2$ is used as the only coarse solver in a classic Parareal method, the method does not converge to the fine serial solution. In the Parareal-HODMD algorithm, when $\mathcal{G}_2$ is run concurrently with $\mathcal{G}_1$, and is used as a correction to $\mathcal{G}_1$ based on the HODMD prediction, the algorithm converges.
 
Figure \ref{fig:visualize_sphere} (a) shows the position of three different tracer points placed on the unit sphere, which are used to track the motion of sphere in the flow, and Figure \ref{fig:visualize_sphere} (b) shows the trajectory of the three colored tracers in (a), computed using the fine solver in serial for $t\in[0,100]$. Figure \ref{fig:visualize_sphere} (c) shows the true relative error $\eta^k$ as the number of iterations increases, obtained using the Parareal-HODMD method and the original Parareal method, which are very close. In Table \ref{tab:runtime_sphere}, we present a comparison of the runtime and accuracy of the two algorithms computed for two different time lengths, $T=10$ and $T=100$. As can be observed from Table \ref{tab:runtime_sphere}, for $T=10$, the Parareal-HODMD algorithm has a runtime that is approximately 50\% of that of the original Parareal algorithm, with a true relative error $\eta^1$ that is just one order of magnitude larger than the original Parareal algorithm. The Parareal-HODMD algorithm achieves a speedup of 23.09, compared to a mere speedup of 12.20 of the original Parareal algorithm. For $T=100$, while the Parareal-HODMD algorithm yields a true relative error $\eta^1$ larger than the original Parareal algorithm after one iteration, for an accuracy requirement of $\eta^k<10^{-5}$, the Parareal-HODMD algorithm can achieve a speedup of 20.76, which approximately doubles that of the original Parareal algorithm, with a speedup of 11.12. Since the same number of cores is used for both $T=10$ and $T=100$, each partitioned sub-interval has larger time length in the case $T=100$. The less ideal accuracy of the Parareal-HODMD solution in the longer-time simulation can be partly attributed to the decreased accuracy of the HODMD prediction performed over the longer period. This can be resolved by increasing the number of cores used for longer-time simulations, as detailed in Section \ref{ssec:scaling}.
\begin{table}[htbp!]
\caption{Runtime and accuracy of the Parareal-HODMD algorithm and the original Parareal algorithm applied to an elastic sphere in shear flow}
\centering
\begin{tabular}{c|c|c|c|c|c|c}
\hline
\multirow{2}{*}{Algorithm} & \multicolumn{3}{|c|}{$T=10$} & \multicolumn{3}{|c}{$T=100$} \\
\hhline{~------} & Runtime (s) & Speedup  & $\eta_1$ & Runtime (s) & Speedup & $\eta_1$\\
\hline
Original Parareal & 868.01 & 12.20 & $6.16\cdot 10^{-8}$ & 8444.15
 & 11.12 & $2.21\cdot 10^{-8}$\\
\hline
Parareal-HODMD & 458.59 & 23.09 & $5.65\cdot10^{-7}$ & 4523.09 & 20.76 & $2.87\cdot 10^{-6}$\\
\hline
\end{tabular}
\label{tab:runtime_sphere}
\end{table}

\subsection{Thin liquid film on a two-dimensional substrate}\label{ssec:film}
In the limit of low Reynolds number, the governing equations for a slowly-varying free surface flow of a viscous liquid coating a solid surface can be reduced to an evolution equation for the film thickness. The lubrication model describing film flow subject to strong surface tension effects on a homogeneous partially wetting solid substrate is a fourth-order nonlinear parabolic differential equation. The evolution of such thin liquid film in two dimensions described by $h(x,y,t)$ is modeled by
\begin{equation}\label{eq:film_2d}
{\partial h\over \partial t} = \nabla \cdot\left(h^3 \nabla \left[\tilde{\Pi}(h) -\nabla^2 h\right]\right),
\end{equation}
where $\tilde{\Pi}(h)$ is the disjoining pressure, which describes the wetting properties of the fluid on the substrate. Here, we consider the simple form $\tilde{\Pi}(h)=A(x,y)\Pi(h)$ where $A(x,y)$ is a function that characterizes the Hamaker constant. $A(x,y)\equiv A$ describes a chemically homogeneous substrate with homogeneous wetting properties. A spatially dependent $A(x,y)$ models a chemically heterogeneous substrate. In this numerical experiment, we consider thin liquid flow on a square heterogenous substrate with a piecewise pattern, described by $A(x,y)$ of the form:
\begin{equation}\label{eq:A_form}
A(x)=
\begin{cases}
A_1 & |x-1/2L|\leq s_1 \ {\rm and} \ |y-1/2L|\leq s_2,\\
A_2& {\rm else}.
\end{cases}
\end{equation}
Here, $L$ is the side length of the square domain, $s_1$ and $s_2$ are the interface of segmentation in $x$ and $y$-direction and $A_i$ are positive constants. Similar forms of $A(x)$ have been previously considered in \cite{2020Steady,2006Rupture}. We normalize $A_i$'s by taking $A_1=1$ and $A_2>A_1$, where the outer region is more hydrophobic than the inner region. Specifically, we consider a disjoining pressure given by a 3-4 inverse power law function which has been previously used in \cite{schwartz1998,oron1999dewetting,oron2001dynamics,glasner03coarsening},
\begin{equation}\lbl{eq:dis34}
\Pi(h)=\frac{\epsilon^2}{h^3}-\frac{\epsilon^3}{h^4},
\end{equation}
where $\epsilon>0$ is a small positive parameter that describes the minimum film thickness. The scaling in \eqref{eq:dis34} yields a finite limit for the effective contact angle of droplet solutions as $\epsilon\to 0$\cite{glasner03coarsening}. 

Fourth-order parabolic equations are stiff, and the restriction on the allowable time step becomes more prohibitive as the size of $\epsilon$ becomes smaller. We apply the Parareal algorithm and the Parareal-HODMD algorithm to solve for the evolution of thin liquid films on a rectangular domain. To solve for the thin film equation, we employ the approximate Newton-ADI scheme \cite{2003ADI}. We briefly describe the numerical approach. To march the solution forward in time, we use the implicit Euler's method, which gives rise to the discretized nonlinear problem at each time step:
\begin{equation}\label{eq:discretized_nonlinear}
F(h^{n})\equiv h^{n}-h^{n-1}+\Delta t\nabla\cdot\left({h^{n}}^3 \nabla \left[\tilde{\Pi}(h^{n}) -\nabla^2 h^{n}\right]\right),
\end{equation}
which needs to be solved for $h^{n}$, the numerical solution of $h(x,y,t_n)$ for $t_n=n\tau$, with $\tau$ denoting the time step size. For spatial discretization, we apply the central finite difference scheme with a five-point stencil. Applying the Newton's method to the nonlinear problem in Equation \eqref{eq:discretized_nonlinear} leads to a linear system to be solved at each iteration:
\begin{equation}\label{eq:jacobi_eq}
J(\tilde{h}^{n+1}_k)(\tilde{h}_{k+1}-\tilde{h}_k)=-F(\tilde{h}_k),\qquad k=0,1,2,\cdots
\end{equation}
where $J$ is the Jacobian of the nonlinear operator $F$. To efficiently solve the large linear system in Equation \eqref{eq:jacobi_eq}, an approximate factorization $J\approx A\equiv L_xL_y$ where $L_x$ and $L_y$ are some linearized operator in $x$ and $y$ dimension respectively is used. With such factorization, the solution to Equation \eqref{eq:jacobi_eq} can be written as a first-order linear system given by 
\begin{align}
L_xw&=-F(\tilde h^{n}),\\
L_yv&=w,\\
\tilde{h}^{n}_{k+1}&=\tilde{h}^{n}_{k}+v.
\end{align}
We note that in matrix form, $L_x$ and $L_y$ are both sparse banded linear systems. We apply LU factorization to solve the two matrix equations. For the stopping criterion inside each Newton's iteration, we compute $A_k e_k =  -F(h_k)$ up to a tolerance $\xi_{Newton}$, i.e.
\begin{equation}\label{eq:newton_iter}
\lVert -A_ke_k + F(h_k)\rVert_2<\xi_{Newton}.
\end{equation}

\begin{figure}[htbp!]
 \begin{subfigure}[b]{0.45\linewidth}
 \centering
\includegraphics[height=1.5in]{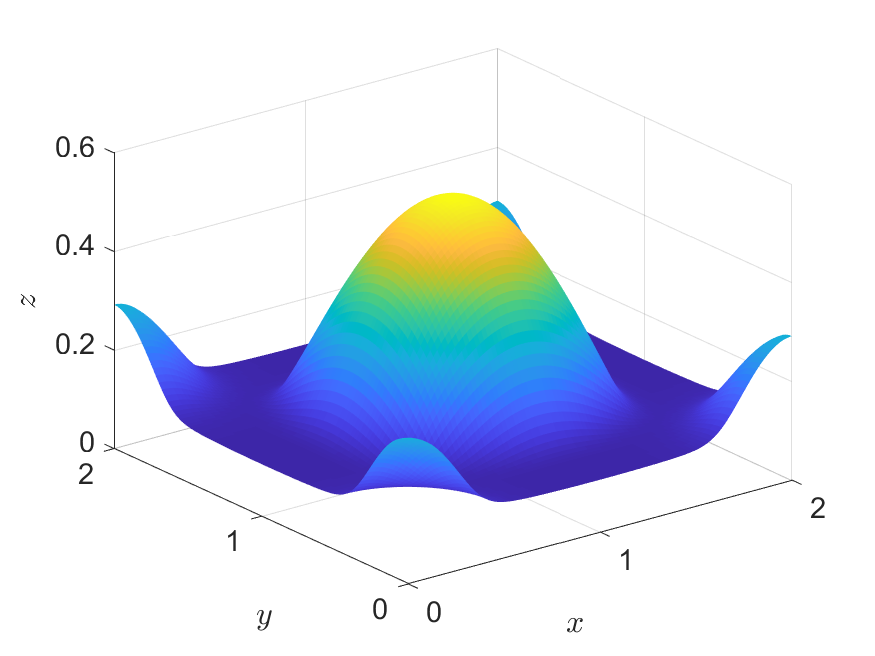}
\caption{}\label{fig:film2}
\end{subfigure}
 \begin{subfigure}[b]{0.45\linewidth}
 \centering
\includegraphics[height=1.5in]{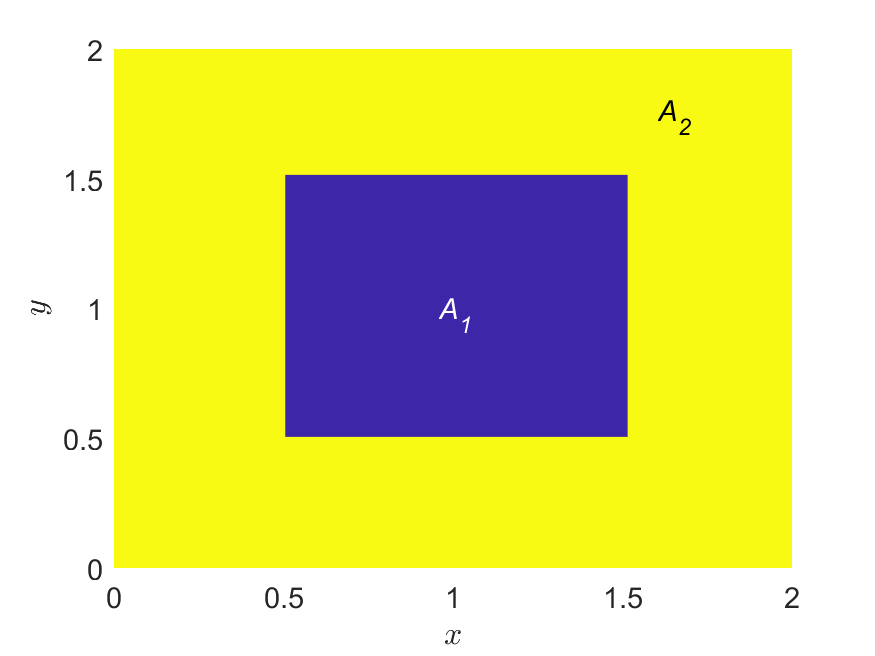}
\caption{}\label{fig:Aplot}
\end{subfigure}
\caption{(a) Thin film at $t=2$ (b) Plot of $A(x,y)$ in the x-y plane with $A_2>A_1$.}
\label{fig:visualize_film}
\end{figure}

In the numerical example, for the pattern of the heterogeneous substrate, we consider $L=2$, $s_1=s_2=1/4L$, $\epsilon=0.1$, $A_1=1$, $A_2=10$. We solve for the evolution of film for time $t\in[0, 2]$ using 10 cores. For the fine solver of both the classic Parareal method and the Parareal-HODMD method, we discretize the domain with $N_x=N_y=100$ points in both $x$ and $y$-direction, resulting in a total of 10,000 points, and thus a Jacobian matrix of size $10,000\times 10,000$. For the time dimension of the fine solver, we employ the implicit Euler's method with $\tau_f=10^{-3}$. To solve for the matrix equation involving the Jacobian matrix, an approximate Newton-ADI scheme is applied where we set $\xi_{Newton}=10^{-5}$ for Newton's iteration in Equation \eqref{eq:newton_iter}. For $\mathcal{G}_2$ of the Parareal-HODMD algorithm, in the time dimension, we use the implicit Euler's method with a time step $\tau_c^{(2)}=10\tau_f=10^{-2}$; we coarsen the spatial grid, using $N_x=N_y=50$ points in the $x$ and $y$-direction, which yields a Jacobian matrix of size $2,500\times 2,500$. Inside $\mathcal{G}_2$, the same tolerance $\xi_{Newton}=10^{-5}$ is used inside of each Newton's iteration. For $\mathcal{G}_1$ of the Parareal-HODMD algorithm, we use the implicit Euler's method with a time step $\tau_c^{(1)}=10\tau_f=10^{-2}$ and spatial grid discretization $N_x=N_y=100$. For the HODMD prediction, we choose $K_t=4$, $l_k=100$, with $N_q=200$, $q_1=q_2=1$. In Figure \ref{fig:visualize_film} (a), we show the fine serial solution at the final time $t=2$. In Figure \ref{fig:visualize_film} (b), we show a plot of the substrate patterning $A(x,y)$, viewed in the x-y plane. Since $A_2>A_1$, the outer rectangular region is more hydrophic, meaning the fluid will be more concentrated in the $A_1$ region as it evolves towards the steady state.

In Table \ref{tab:runtime_film}, we compare the runtime and accuracy of the solution to the thin film equation obtained using the original Parareal algorithm and the Parareal-HODMD algorithm. Specifically, we perform two versions of the original Parareal algorithm, where in one experiment, we take the coarse solver to be the same solver as $\mathcal{G}_1$ used in the Parareal-HODMD method, i.e. with $N_x=N_y=100$. In another experiment, we take the coarse solver to be the same solver as $\mathcal{G}_2$ used in the Parareal-HODMD method, i.e. $N_x=N_y=50$. In both experiments, we set $\tau_c=10\cdot\tau_f=10^{-2}$. We show the results of the two simulations in the first two rows of Table \ref{tab:runtime_film}. We see that it takes $k=3$ iterations of the coarser version of the algorithm to produce a solution of approximately the same order of accuracy as that of the finer version, where $k=1$. In the last row of Table \ref{tab:runtime_film}, we show the results of the Parareal-HODMD algorithm. For a solution accuracy $\eta^k<10^{-4}$, compared to the more accurate original Parareal algorithm which uses $\mathcal{G}_1$ as the coarse solver, the Parareal-HODMD algorithm has a runtime that is approximately 75\% of that of the original Parareal algorithm. Compared to the less accurate original Parareal algorithm that uses $\mathcal{G}_2$ as the coarse solver, the Parareal-HODMD has a runtime that is approximately 80\% of that of the original Parareal algorithm. Both results show that the Parareal-HODMD equipped with $\mathcal{G}_1$ and $\mathcal{G}_2$ has a greater speedup than the original Parareal equipped with $\mathcal{G}_1$ or $\mathcal{G}_2$. Compared to the numerical example of the rod microswimmer, the efficiency improvement given by the Parareal-HODMD relative to the original Parareal is less prominent. First, this is due to the larger number of spatial grid points and the larger matrix size involved, which results in an increased communication cost between cores and an increased computation cost of the SVD involved in the HODMD prediction. Second, unlike the swimmer example which uses an explicit time marching scheme in both of the fine and coarse solvers, and thus has a fixed computational complexity at each time step, the thin film example employs an implicit time scheme and uses an iterative method to solve the nonlinear problem. While using a coarser mesh has a lower cost per Newton's iteration, the reduced solution accuracy may increase the number of iterations required, resulting in a more complicated interplay between grid resolution and total runtime.

\begin{table}[htbp!]
\caption{Runtime and accuracy of the Parareal-HODMD algorithm and the original Parareal algorithm applied to the thin film equation}
\centering
\begin{tabular}{c|c|c|c|c}
\hline
Algorithm & Coarse solver parameters & Runtime (s) & Speedup & $\eta^k$ \\
\hline
\multirow{2}{*}{Original Parareal} & $N_x=N_y=100$, $k=1$ & 4638.77 & 1.46 & $1.43\cdot 10^{-5}$ \\
\hhline{~----}& $N_x=N_y=50$, $k=3$  &  4377.17 & 1.55 & $2.11\cdot 10^{-5}$\\
\hline
Parareal-HODMD & $K_t=4$, $r_q=1$, $l_1=10$, $k=1$ & $ 3462.14$ & 1.95 & $6.77\cdot 10^{-5}$\\
\hline
\end{tabular}
\label{tab:runtime_film}
\end{table}

\subsection{Scaling of the algorithm}\label{ssec:scaling}
\begin{figure}[htbp!]
\centering
\includegraphics[height=2in]{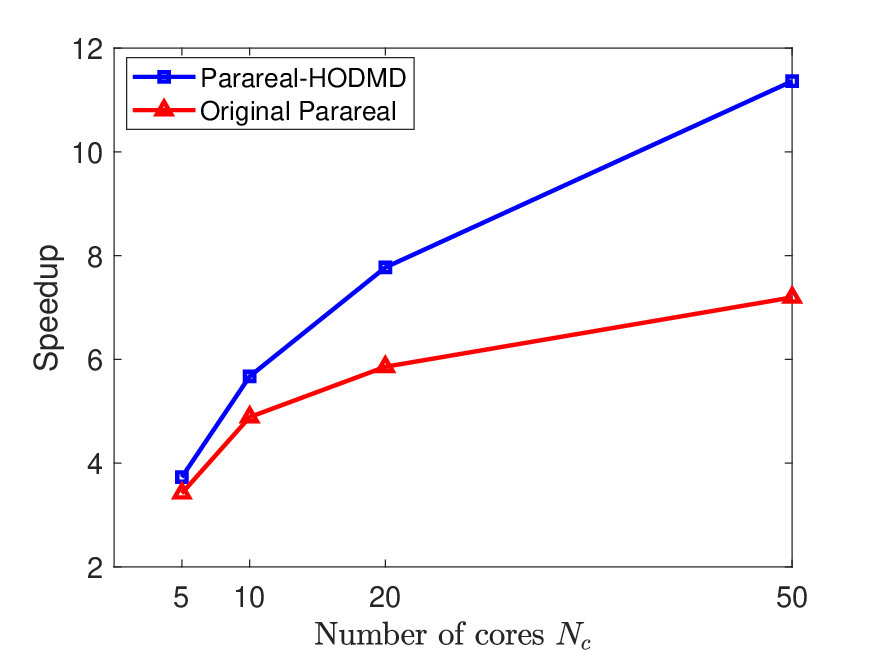}
\caption{Strong scaling of Parareal-HODMD and the original Parareal algorithm}
\label{fig:scaling1}
\end{figure}

In this section, we present the strong scaling of the Parareal-HODMD algorithm and the original Parareal algorithm. We solve for the motion of a flagellar microswimmer for $t\in[0,1]$ using both algorithms with $\nc=5, ~10, ~20, ~50$ cores. The speedup of the Parareal method is greatly influenced by the number of iterations or the desired error tolerance. We compare the scaling results of the algorithm for solution accuracy $\eta^k<\vartheta=10^{-5}$ where one iteration is sufficient for both algorithms. In the Parareal-HODMD algorithm, for $\nc=5, 10, 20, 50$, we use $K_t=0.6\nc$, $l_1=0.48N_q$, $q_2=50, 50, 5, 5$, and $d_2=12, 12, 12, 5$ respectively. For all of the four core numbers, we set $\tau_c^{(1)}=\tau_c^{(2)}=8\tau_f$, $\sigma_c^{(1)}=\sigma_f$, $\sigma_c^{(2)}=6\sigma_f$, $d_1=10$ and $q_1=50$. In the original Parareal algorithm, we set $\tau_c=8\tau_f$ and $\sigma_c=\sigma_f$. We plot the strong scaling graph for the two algorithms when 5, 10, 20 and 50 cores are used in Figure \ref{fig:scaling1}. We show the speedup in Table \ref{tab:scaling_tab}.

With the DMD parameters chosen above, the Parareal-HODMD method yields solution of approximately the same accuracy after one iteration for $\nc=10, 20, 50$. We observe an increasing gap between the speedup of the two algorithm as the number of cores increases. As HODMD is based on a linearized approximation, its precision deterioates for longer time prediction. For fixed $T$, when $\nc$ is smaller, $\Delta T=T/\nc$ is larger. For larger $\Delta T$, in the serial correction stage of the algorithm where HODMD is applied per subinterval, a larger number of snapshots is needed to construct the snapshot matrix, since the prediction needs to be made for longer time. As $\nc$ increases, $\Delta T$ becomes smaller. When HODMD is applied to obtain shorter time prediction, the snapshot matrix is smaller and the prediction accuracy is higher. This also implies that the Parareal-HODMD method is more efficient and accurate for larger core numbers, and can be particularly effective for breaking the speedup bottleneck and saturation observed in many Parareal simulations when the number of cores is large. In the example of the rod-like swimmer, we find that for $\nc=5$, the speedup of the original Parareal and Parareal-HODMD is approximatley 3.42 and 3.73 respectively, which is quite close. As $\nc$ increases, the Parareal-HODMD algorithm yields an increasingly more remarkable improvement in the speedup compared to the original Parareal algorithm. We note that for $\nc=50$, the Parareal-HODMD algorithm is $3.29\cdot 10^{4}$ seconds faster than the original Parareal algorithm, which is approximately 9.14 hours of the runtime difference.

\begin{table}[htbp!]
\caption{Speedup of the strong scaling experiment of Parareal-HODMD and the original Parareal algorithm}
\centering
\begin{tabular}{c|c|c|c|c}
\hline
{$\nc$} & 5 & 10  & 20 & 50 \\
\hline
{Original Parareal} & 3.42 & 4.89 & 5.86 & 7.20\\
\hline
{Parareal-HODMD} & 3.73 & 5.67 & 7.77 & 11.36\\
\hline
\end{tabular}
\label{tab:scaling_tab}
\end{table}

\section{Conclusion}\label{sec:conclude}
To improve the computation efficiency of long-time numerical simulations often require the application of some parallel-in-time approach like the Parareal algorithm. For many stiff problems, using large time steps is often prohibited, resulting in extremely expensive computation of the coarse propagator of the Parareal method. Hence, the high cost and sequential nature of the coarse solver can significantly limit the parallel efficiency of the Parareal algorithm. In this work, we propose a novel Parareal algorithm named Parareal-HODMD algorithm by constructing the coarse solution using the solution from part of a time interval based on High-Order Dynamic Mode Decomposition. Instead of computing only one coarse solver in the serial stage, the proposed Parareal-HODMD algorithm launches two coarse solvers in parallel, with one core solving the equation with higher accuracy for only part of the time interval and the other core solving the equation with lower accuracy for the entire time interval. The algorithm computes its coarse solution by calculating the difference between the two coarse solutions on part of the time interval and predicting the more accurate coarse solver at a future time by applying HODMD to the difference snapshot matrix.

By introducing a second coarse solver and HODMD, the Parareal-HODMD algorithm lowers the cost of sequential computation by reducing the number of time steps in the solver. The use of HODMD allows for a more flexible adjustment of the cost of the coarse solver based on the desired error tolerance, and thus allows for improvement in the speedup of simulations where relaxing the time step is difficult. We demonstrate through various numerical experiments, for not too stringent stopping criterions, when the parameters are chosen appropriately, the proposed algorithm can provide a significant acceleration of some long-time simulations of complicated fluid problems that are hard to accelerate using the classic Parareal method.

Our future work includes the two following directions. First, in this work, the algorithm is applied to only simulations of limited spatial scale. The cost of SVD can increase rapidly as the spatial scale of the problem increases. It would be valuable to consider variants of DMD methods with more time and memory efficient SVD techniques, such as randomized SVD \cite{HalkoRSVD}, and apply the algorithm to larger-scale fluid simulations of both increased spatial and time complexity. Second, only two cores are used in the computation of the coarse solver in the proposed Parareal-HODMD algorithm, while $\nc - 2$ cores are idle during the same time. For problems with large spatial scale, the computations involved in the DMD prediction such as the SVD process can be further parallelized using the idle cores to improve efficiency.In addition, previous work has considered the use of idle cores by means of parameter extrapolation for improving the accuracy of the coarse solver\cite{liu2022parallel}. Incorporating such extrapolation techniques into the Parareal-HODMD algorithm using the idle cores may also further enhance the accuracy, efficiency and flexibility of the algorithm. 

\section*{Acknowledgements}
This work is supported by National Natural Science Foundation of China (Grant Number 12301547) and the Fundamental Research Funds for the Central Universities (Grant Number BLX202245). The computation of this work is supported by the high-performance computing platform of Beijing Forestry University.

\section*{Data availability}
The code for the numerical examples presented in this work is available on Github. Link to the repository: \url{https://github.com/weifan-liu/parareal_hodmd}.

\bibliographystyle{unsrt}  
\bibliography{main}

\end{document}